\documentclass[a4paper,10pt]{article}

\usepackage{geometry}

\usepackage[T1]{fontenc}
\usepackage[utf8]{inputenc}

\usepackage{amssymb, amsthm, amsmath}

\usepackage{authblk}
\usepackage[hidelinks]{hyperref} 
\usepackage{url}
\usepackage{color}

\usepackage{tikz}
\usetikzlibrary{calc}

\theoremstyle{plain}
\newtheorem{theorem}{Theorem}[section]
\newtheorem{lemma}[theorem]{Lemma}
\newtheorem{proposition}[theorem]{Proposition}
\newtheorem{corollary}[theorem]{Corollary}
\newtheorem{conjecture}[theorem]{Conjecture} 

\theoremstyle{definition}
\newtheorem{definition}[theorem]{Definition} 
\newtheorem{remark}[theorem]{Remark}
\newtheorem{example}[theorem]{Example}

\newcommand{\wt}{{\rm wt}}
\newcommand{\dd}{{\rm d}}
\newcommand{\C}{\mathcal{C}}
\newcommand{\M}{\mathcal{M}}
\newcommand{\PP}{\mathcal{P}}
\newcommand{\LL}{\mathcal{L}}
\newcommand{\K}{\mathcal{K}}
\newcommand{\V}{\mathcal{V}}
\newcommand{\A}{\mathcal{A}}
\newcommand{\F}{\mathbb{F}}
\newcommand{\N}{\mathbb{N}}
\newcommand{\Z}{\mathbb{Z}}

\newcommand{\GL}{{\rm GL}}
\newcommand{\diag}{{\rm diag}}

\newcommand{\cc}{\mathbf{c}}
\newcommand{\xx}{\mathbf{x}}

\newcommand{\aaa}{\mathbf{a}}
\newcommand{\bb}{\mathbf{b}}

\newcommand{\nn}{\mathbf{n}}
\newcommand{\bbeta}{\boldsymbol\beta}
\newcommand{\aalpha}{\boldsymbol\alpha}
\newcommand{\ggamma}{\boldsymbol\gamma}
\newcommand{\col}{\mathrm{Col}}
\newcommand{\row}{\mathrm{Row}}
\newcommand{\supp}{\mathrm{Supp}}
\newcommand{\rk}{\mathrm{Rk}}
\newcommand{\qmultinom}[2]{\genfrac{[}{]}{0pt}{}{#1}{#2}_q}

\title{The ordered sum-rank metric and linearized\\multiplicity codes: NRT meets sum-rank}
\author{Umberto Mart{\'i}nez-Pe\~{n}as \thanks{umberto.martinez@uva.es}}
\affil{IMUVa-Mathematics Research Institute,\\University of Valladolid, Spain}

\date{}

\begin{document}

\maketitle

\begin{abstract}
In this work, we introduce linearized multiplicity codes, which are a common generalization of linearized Reed--Solomon codes and univariate multiplicity codes. We also introduce the ordered sum-rank metric, which is a common generalization of the sum-rank metric and the Niederreiter-Rosenbloom-Tsfasman (NRT) metric (or ordered Hamming metric). We provide a Singleton bound for the ordered sum-rank metric and show that linearized multiplicity codes attain it. That is, they are maximum ordered sum-rank distance (MOSRD) codes. We introduce the support lattice for the ordered sum-rank metric and study duality, support distributions and weight distributions of MOSRD codes. We also show that MOSRD codes are ordered MDS, dually quasi MSRD and, under some divisibility conditions, MSRD (and thus MDS). For linear MOSRD codes, we give lower bounds on the field extension degree by the number of matrix columns, and upper bounds on the block length by the base field size. We conclude by showing that linearized multiplicity codes attain such bounds exactly in almost all cases, and the same growth order in all cases. 

\end{abstract}

\begingroup
  \renewcommand{\baselinestretch}{0.75}\normalsize
  \tableofcontents
\endgroup

\section{Introduction} \label{sec_intro}

Reed--Solomon codes \cite{reed-solomon} are the best known MDS codes, that is, codes that attain the Singleton bound for the Hamming metric. Their length can attain the field size plus one, which is conjectured to be maximum (this is the well-known MDS conjecture, proven for prime fields \cite{ball}).

Reed--Solomon codes are constructed by evaluating polynomials of restricted degree over a finite field. Later, they were extended to include also their derivatives up to a certain order. This results in univariate multiplicity codes \cite{multiplicity, nrt}, also known as hyperderivative Reed--Solomon codes \cite{hyperderivative}. They attain the Singleton bound for the Niederreiter-Rosenbloom-Tsfasman (NRT) metric \cite{niederreiter, nrt}, i.e., they are ordered MDS \cite{barg_park} (since the NRT metric is also known as the ordered Hamming metric \cite{barg_poset}). Interestingly, by adding the point at infinity, multiplicity codes also have length equal to the field size plus one, which is already proven (for some dimensions) to be maximum for all fields for multiplicity larger than one \cite[Cor. 4.21]{Niederreiter1992} (see also \cite[Eq. (3.27)]{skriganov}).

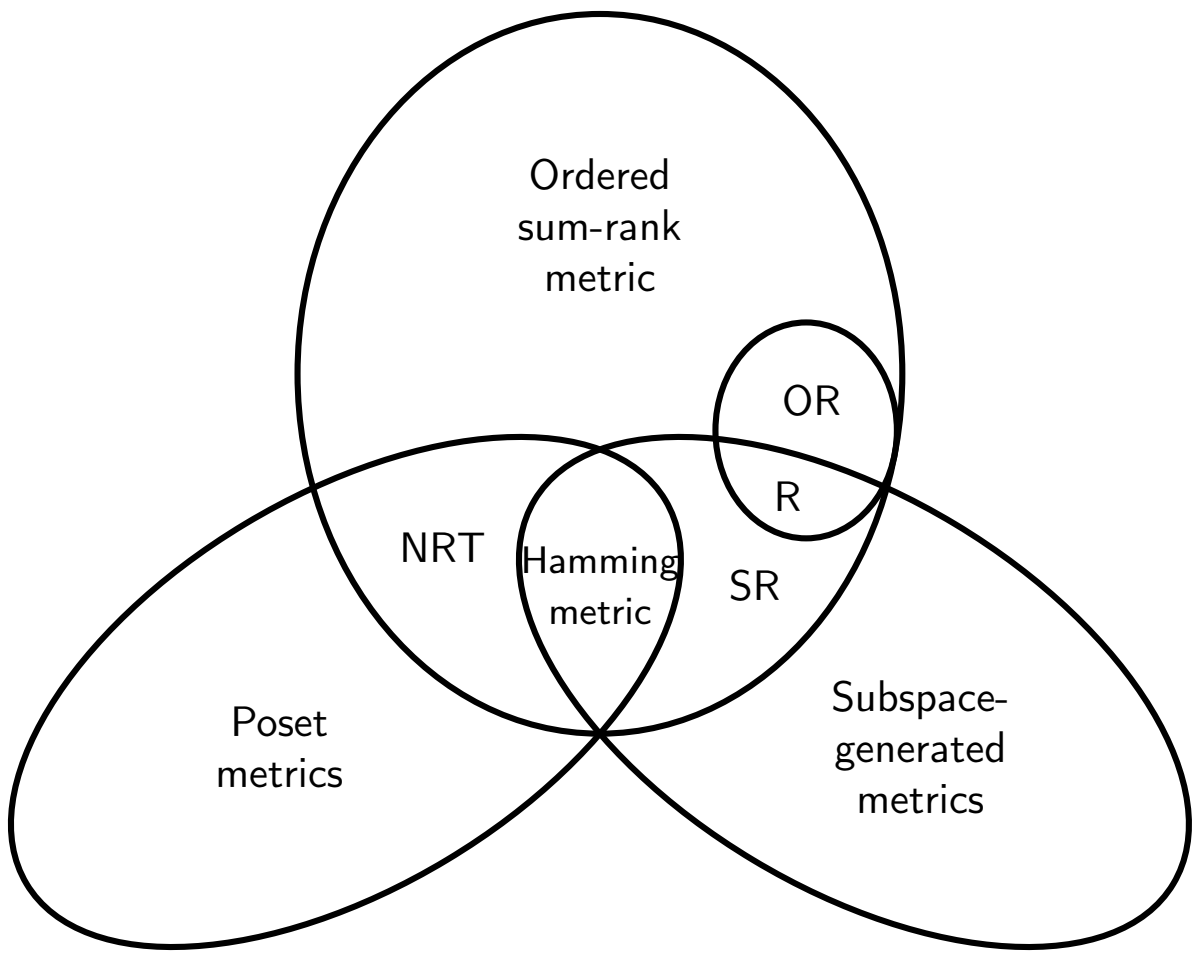
\begin{figure}[t!]
    \centering
    
\begin{tikzpicture}[
	scale=0.8,
    transform shape,
    line width=1.15pt,
    font=\sffamily
]

\draw[
    rotate around={31:(-2.10,-1.08)}
]
    (-2.10,-1.08)
    ellipse [x radius=3.08, y radius=1.62];

\draw[
    rotate around={-31:(2.10,-1.08)}
]
    (2.10,-1.08)
    ellipse [x radius=3.08, y radius=1.62];

\def\OSRcy{1.55}
\def\OSRrx{2.50}
\def\OSRry{2.9746211683}

\draw
    (0,\OSRcy)
    ellipse [x radius=\OSRrx, y radius=\OSRry];

\draw
    (1.70515,1.08225)
    ellipse [x radius=0.75000, y radius=0.89239];

\node[align=center] at (0,2.78) {Ordered\\sum-rank\\metric};

\node at (-1.3,0.12) {NRT};
\node at ( 1.28,-0.2) {SR};
\node[align=center] at (0,-0.2) {{\small Hamming}\\{\small metric}};

\node[align=center] at (-2.65,-1.53) {Poset\\metrics};
\node[align=center] at ( 2.65,-1.53) {Subspace-\\generated\\metrics};

\node at (1.75,1.33) {OR};
\node at (1.55,0.54) {R};

\begin{scope}[xshift=8.2cm]

\draw
    (0,\OSRcy)
    ellipse [x radius=\OSRrx, y radius=\OSRry];

\begin{scope}
    \clip (0,\OSRcy) ellipse [x radius=\OSRrx, y radius=\OSRry];

    \draw[
        rotate around={31:(-2.10,-1.08)}
    ]
        (-2.10,-1.08)
        ellipse [x radius=3.08, y radius=1.62];

    \draw[
        rotate around={-31:(2.10,-1.08)}
    ]
        (2.10,-1.08)
        ellipse [x radius=3.08, y radius=1.62];
\end{scope}

\draw
    (1.70515,1.08225)
    ellipse [x radius=0.75000, y radius=0.89239];

\node[align=center] at (0,2.78) {Linearized\\multiplicity\\codes};

\node[align=center] at (-1.42,0.2) {{\small Multiplicity}\\{\small codes}};
\node[align=center] at ( 1.15,-0.25) {LRS\\codes};
\node[align=center] at (0,0) {RS\\codes};

\node[align=center] at (1.55,0.6) {Gab};

\end{scope}

\end{tikzpicture}

\caption{Venn diagrams illustrating, on the left, the relationship among the poset metrics, subspace-generated metrics, the NRT metric, the sum-rank (SR) metric, the ordered sum-rank metric (new), the ordered rank (OR) metric (new), the rank (R) metric and the Hamming metric; and on the right, the codes that attain the corresponding Singleton bound: linearized multiplicity codes (new), multiplicity codes, linearized Reed--Solomon (LRS) codes, Gabidulin (Gab) codes and Reed--Solomon (RS) codes.}
    \label{fig_metrics}
\end{figure}

In another line of research, linearized Reed--Solomon codes, introduced in \cite{linearizedRS}, extend Reed--Solomon codes by considering linearized evaluations of skew polynomials \cite{ore}. Linearized Reed--Solomon codes attain the Singleton bound for the sum-rank metric \cite{space-time-kumar, multishot}, and are therefore maximum sum-rank distance (MSRD). For square matrices of size at least two, their block length can be the field size minus one, which is conjectured to be maximum in most cases \cite{astore}. Linearized Reed--Solomon codes are a common generalization of Reed--Solomon codes and Gabidulin codes \cite{gabidulin}, and the sum-rank metric is a common generalization of the Hamming metric and the rank metric. See also the monograph \cite{fnt}.

Later, the NRT metric was generalized to poset metrics \cite{poset}, and the rank metric was generalized to subspace-generated metrics \cite{gabidulin_metrics}, which also include the sum-rank metric. However, the intersection between the families of poset metrics and subspace-generated metrics is exactly the Hamming metric, see \cite[Sec. 3]{firer_poset} and Figure \ref{fig_metrics}.

In this work, we introduce linearized multiplicity codes, which are a common generalization of multiplicity codes and linearized Reed--Solomon codes. We then introduce the ordered sum-rank metric, which is a common generalization of the NRT metric and the sum-rank metric. We show that linearized multiplicity codes attain the Singleton bound for the ordered sum-rank metric, that is, they are maximum ordered sum-rank distance (MOSRD) codes. We also show that, in almost all cases, they attain the largest block length relative to the base field size and the smallest field extension degree relative to the column numbers. Hence they are extremal MOSRD codes. Interestingly, the intersection between the ordered sum-rank metric and the family of poset metrics is the NRT metric, and the intersection between the ordered sum-rank metric and the family of subspace-generated metrics is the sum-rank metric. See Figure \ref{fig_metrics}.

The aim of this work is foundational. We deliberately leave applications for future work. The NRT metric has applications beyond classical error correction. Its close connections with ordered orthogonal arrays and digital nets provide a coding-theoretic approach to low-discrepancy point sets and quasi-Monte Carlo methods \cite{martin,niederreiter,Niederreiter1992,skriganov}. It also arises in models of information transmission with ordered interference \cite{barg_park,nrt}. On the other hand, the sum-rank metric has found applications in reliable and secure multishot network coding \cite{secure-multishot,multishot}, maximally recoverable locally repairable codes for distributed storage \cite{universal-lrc}, and rate-diversity optimal multiblock space-time coding \cite{space-time-kumar,mohannad}. The structure of the ordered sum-rank metric suggests some plausible applications. For instance, one could assign different multiplicities to different packets in a multishot linearly coded network \cite{secure-multishot} in order to guarantee a stronger error-correction capability or stronger information-theoretical privacy to certain packets (i.e., unequal protection). 

The contributions and organization of the manuscript are as follows. In Section \ref{sec_lin_multi}, we revisit results on multiplicities of zeros of skew polynomials \cite{skew-multi} and we introduce linearized derivatives and linearized multiplicity codes. In Section \ref{sec_ordered_SR_metric}, we introduce the ordered sum-rank metric and its Singleton bound, and show that linearized multiplicity codes attain it in all cases, i.e., they are MOSRD and, in particular, they are ordered MDS. In Section \ref{sec_supports}, we introduce and study the ordered sum-rank support lattice and prove that the dual of a linear MOSRD code is again MOSRD (for the opposite order). We conclude that section by showing that MOSRD implies dually quasi MSRD \cite{qmsrd} and, under certain divisibility conditions, MSRD (thus also MDS). In Section \ref{sec_weight}, we compute the support and weight distributions of MOSRD codes. In Section \ref{sec_upper_bounds_m}, we provide lower bounds on the extension degree (of the field of linearity) by the number of matrix columns. Finally, in Section \ref{sec_upper_bounds_ell}, we provide upper bounds on the block length by the field size. We show that linearized multiplicity codes attain such bounds exactly in almost all cases and the same growth order in all cases. \\ 

\textbf{Notation:} Denote $ \mathbb{N} = \{ 0,1,2, \ldots \} $ and $ \mathbb{Z}_+ = \{ 1,2,3, \ldots \} $. For $ m,n \in \mathbb{Z} $ we denote $ [m,n] = \{ m,m+1, \ldots, n \} $ (if $ m \leq n $) and $ [n] = [1,n] $ (if $ n \geq 1 $). We fix a prime power $ q $ and denote by $ \F_q $ the finite field with $ q $ elements. For $ m,n \in \Z_+ $, we denote by $ \F_q^{m \times n} $ the space of $ m \times n $ matrices over $ \F_q $, and we define $ \F_q^n = \F_q^{1 \times n} $. For elements $ a,b $ in some set, $ \delta_{a,b} $ denotes the Kronecker delta ($ \delta_{a,b} = 1 $ if $ a=b $, and $ \delta_{a,b}= 0 $ otherwise).

\section{Linearized derivatives and multiplicity codes} \label{sec_lin_multi}

In this section, we revisit some known results on the zeros of skew polynomials, counted with multiplicities, and introduce the new concepts of linearized Hasse derivatives and linearized multiplicity codes.

Let $ m \in \mathbb{Z}_+ $ and define $ \sigma : \mathbb{F}_{q^m} \longrightarrow \mathbb{F}_{q^m} $ by $ \sigma(a) = a^q $, for all $ a \in \mathbb{F}_{q^m} $. The ring of skew polynomials $ \mathbb{F}_{q^m}[x;\sigma] $ was defined by Ore \cite{ore} as the classical ring of polynomials, but with a product given by $ x^i x^j = x^{i+j} $, for $ i,j \in \mathbb{N} $, and
$$ xa = \sigma(a)x, $$
for $ a \in \mathbb{F}_{q^m} $. This ring is a (generally non-commutative) left and right Euclidean domain, and recovers the classical polynomial ring $ \F_q[x] $ when $ m=1 $. Furthermore, defining the degree of a skew polynomial as usual, we have that $ \deg(FG) = \deg(F) + \deg(G) $, for all $ F,G \in \mathbb{F}_{q^m}[x;\sigma] $. 

Using right Euclidean division, Lam and Leroy defined the following arithmetic evaluation of skew polynomials (see \cite[Th. 3]{lam} and \cite[Lemma 2.4]{lam-leroy}).

\begin{definition} [\textbf{Arithmetic evaluation} \cite{lam, lam-leroy}]
Given $ F \in \mathbb{F}_{q^m}[x;\sigma] $ and $ a \in \mathbb{F}_{q^m} $, we define the evaluation of $ F $ in $ a $ as the unique $ F(a) \in \mathbb{F}_{q^m} $ such that there exists $ G \in \mathbb{F}_{q^m}[x;\sigma] $ satisfying 
$$ F = G\cdot (x-a) + F(a) . $$
\end{definition}

We will need the concept of conjugacy, also provided by Lam and Leroy, see \cite[p. 198]{lam} and \cite[Eq. (2.5)]{lam-leroy}.

\begin{definition} [\textbf{Conjugacy} \cite{lam, lam-leroy}] \label{def_ev_arithmetic}
Given $ a, \beta \in \mathbb{F}_{q^m} $ with $ \beta \neq 0 $, we define 
$$ a^\beta = \sigma(\beta)a\beta^{-1} = \beta^{q-1}a . $$
We say that $ a,b \in \mathbb{F}_{q^m} $ are conjugate if there exists $ \beta \in \mathbb{F}_{q^m}^* $ such that $ b = a^\beta $.
\end{definition}

Conjugacy defines an equivalence relation in $ \mathbb{F}_{q^m} $, with $ q-1 $ non-zero conjugacy classes of size $ \frac{q^m-1}{q-1} $ and the zero conjugacy class (with only the element $ 0 $), see \cite[Sec. 2.4]{fnt}. Using conjugacy, we may establish the rule of arithmetic evaluation for a product, given by Lam and Leroy in \cite[Th. 2]{lam} and \cite[Th. 2.7]{lam-leroy}.

\begin{lemma}[\textbf{Product rule} \cite{lam, lam-leroy}] \label{lemma_product_rule}
Let $ F,G \in \mathbb{F}_{q^m}[x;\sigma] $ and $ a \in \mathbb{F}_{q^m} $. If $ G(a) = 0 $, then $ (FG)(a) = 0 $, and if $ \beta = G(a) \neq 0 $, then 
$$ (FG)(a) = F(a^\beta)G(a). $$
\end{lemma}

Interestingly, Definition \ref{def_ev_arithmetic} does not recover the classical evaluation of linearized polynomials. It is recovered using instead the following type of evaluation, first considered in \cite[Def. 3.1]{hilbert90} and \cite[Eq. (2.7)]{leroy-pol}.

\begin{definition} [\textbf{Linearized evaluation} \cite{hilbert90, leroy-pol}] \label{def_ev_linearized}
For $ a \in \mathbb{F}_{q^m} $ and $ i \in \mathbb{N} $, define 
$$ N_i(a) = \sigma^{i-1}(a) \cdots \sigma(a)a = a^{q^{i-1}+\cdots+q+1} = a^{\frac{q^i -1}{q-1}}. $$
Given $ F = \sum_{i=0}^d f_i x^i \in \mathbb{F}_{q^m}[x;\sigma] $ and $ a,\beta \in \mathbb{F}_{q^m} $, we define
$$ F_a(\beta) = \sum_{i=0}^d f_i N_i(a) \sigma^i (\beta) = \sum_{i=0}^d f_i a^{\frac{q^i-1}{q-1}} \beta^{q^i} . $$
\end{definition}

Clearly, $ F_a : \mathbb{F}_{q^m} \longrightarrow \mathbb{F}_{q^m} $ is an $ \mathbb{F}_q $-linear map. Moreover, when $ a=1 $,
$$ F_1(\beta) = f_0 \beta + f_1 \beta^q + f_2 \beta^{q^2} + \cdots + f_d \beta^{q^d}, $$
which is the classical evaluation of linearized polynomials. In general, we can connect the evaluations from Definition \ref{def_ev_arithmetic} and \ref{def_ev_linearized} as follows. The following result is \cite[Lemma 1(2)]{lam} and \cite[Prop. 2.9(1)]{lam-leroy}.

\begin{proposition} [\textbf{Connecting evaluations} \cite{lam, lam-leroy}] \label{prop_connection_evals}
For $ F \in \mathbb{F}_{q^m}[x;\sigma] $ and $ a,\beta \in \mathbb{F}_{q^m} $ with $ \beta \neq 0 $,
$$ F_a(\beta) = F(a^\beta) \beta. $$
\end{proposition}

In the classical case ($ m = 1 $), we recover classical evaluations and the previous proposition reads $ F_a(1) = F(a^1) \cdot 1 = F(a) $ (and in general, $ F_a(\beta) = \beta F(a) $).

Notice also that, given $ a,\beta \in \mathbb{F}_{q^m} $ with $ \beta \neq 0 $, then the previous proposition implies that $ F(a^\beta) = 0 $ if, and only if, $ F_a(\beta) = 0 $. Thus we can use either notion of evaluation to count zeros. Definition \ref{def_ev_linearized} gives linear maps, thus the corresponding zeros form linear subspaces. However, Definition \ref{def_ev_arithmetic} gives a natural way to count zeros, as we next show. The following definition is given in the paragraphs before \cite[Lemma 5]{lam} and \cite[Th. 5.1]{lam-leroy}.

\begin{definition} [\textbf{Minimal skew polynomial} \cite{lam, lam-leroy}]
Given $ \Omega = \{ b_1, b_2, \ldots, b_n \} \subseteq \mathbb{F}_{q^m} $, we define their minimal skew polynomial as
$$ F_\Omega = {\rm llcm} \{ x-b_1, x-b_2, \ldots, x-b_n \} \in \F_{q^m}[x;\sigma] $$
(the least left common multiple, which we define as monic). 
\end{definition}

$ F_\Omega $ can be equivalently defined as the unique monic skew polynomial of minimal degree that vanishes in $ \Omega $ according to arithmetic evaluation (Definition \ref{def_ev_arithmetic}). The next theorem combines \cite[Prop. 6, Lemma 12, Th. 23(1)]{lam} and \cite[Th. 4.5]{lam-leroy}. See also \cite[Ch. 2]{fnt}.

\begin{theorem} [\textbf{P-independence} \cite{lam, lam-leroy}] \label{th_P-independence}
For any nonempty $ \Omega = \{ b_1,b_2,  \ldots, b_n \} \subseteq \mathbb{F}_{q^m} $, of size $ n = |\Omega| $, we have that
$$ \deg(F_\Omega) \leq |\Omega|. $$
Furthermore, the following are equivalent:
\begin{enumerate}
\item
$ \deg(F_\Omega) = | \Omega | $.
\item
For any nonzero $ F \in \mathbb{F}_{q^m}[x;\sigma] $ with $ F(b_1) = F(b_2) = \ldots = F(b_n) = 0 $, we have that
$$ |\Omega | \leq \deg(F) . $$
\item
(If $ \Omega \subseteq \F_{q^m}^* $) If we choose pairwise nonconjugate $ a_1, a_2, \ldots , a_\ell \in \mathbb{F}_{q^m}^* $, $ n_i \in \mathbb{Z}_+ $ and distinct $ \beta_{i,1}, \beta_{i,2}, \ldots, $ $ \beta_{i,n_i} \in \mathbb{F}_{q^m}^* $, for $ i \in [\ell] $, such that $ |\Omega| = n_1 + n_2 + \cdots + n_\ell $ and
$$ \Omega = \bigcup_{i=1}^\ell \left\lbrace a_i^{\beta_{i,1}}, a_i^{\beta_{i,2}}, \ldots , a_i^{\beta_{i,n_i}} \right\rbrace , $$
then $ \beta_{i,1}, \beta_{i,2}, \ldots, \beta_{i,n_i} \in \mathbb{F}_{q^m}^* $ are $ \mathbb{F}_q $-linearly independent, for $ i \in [\ell] $.\footnote{There can be $ \mathbb{F}_q $-linear dependencies among the beta elements across different blocks, i.e., for different $ i,j \in [\ell] $.}
\end{enumerate}
We say that $ \Omega $ is a P-independent set if it satisfies any of the above properties.
\end{theorem}

Theorem \ref{th_P-independence} essentially says that, in order to recover classical results (e.g., bounding the number of zeros by the degree), we need to consider only P-independent sets. 

However, in this work, we will need to count zeros with multiplicities. The naive idea of defining $ a \in \mathbb{F}_{q^m} $ as a root of $ F \in \mathbb{F}_{q^m}[x;\sigma] $ of multiplicity $ r \in \mathbb{Z}_+ $, if $ (x-a)^r $ divides $ F $ on the right, does not allow for results such as Theorem \ref{th_P-independence}. See \cite[Ex. 21]{skew-multi}. Instead, we will consider the following alternative definition of multiplicities. The two following definitions are \cite[Def. 26]{skew-multi} and \cite[Def. 28]{skew-multi}, respectively.

\begin{definition} [\textbf{Multiplicity sequences} \cite{skew-multi}]
For $ \mathbf{a} = (a_1,a_2, \ldots, a_r) \in \mathbb{F}_{q^m}^r $, we define
$$ P_\mathbf{a} = (x-a_r) \cdots (x-a_2)(x-a_1) \in \mathbb{F}_{q^m}[x;\sigma]. $$
We say that $ \mathbf{a} $ is a multiplicity sequence if $ a_1 $ is the only zero of $ P_\mathbf{a} $ on the right, i.e., according to Definition \ref{def_ev_arithmetic}.
\end{definition}

\begin{definition} [\textbf{Multiplicity} \cite{skew-multi}] \label{def_multiplicity}
Given $ F \in \mathbb{F}_{q^m}[x;\sigma] $, $ r \in \mathbb{Z}_+ $, and a multiplicity sequence $ \mathbf{a} = (a_1, a_2, \ldots, a_r) \in \mathbb{F}^r $, we say that $ a_1 $ is a zero of $ F $ of multiplicity $ r $ (via $ \mathbf{a} $) if $ P_{\mathbf{a}} $ divides $ F $ on the right. 
\end{definition}

In \cite[Ex. 21]{skew-multi}, it was shown that if $ m > 1 $, then $ (a,a, \ldots, a) \in \mathbb{F}_{q^m} $ is not always a multiplicity sequence. Conversely, if $ m > 1 $, there may exist multiplicity sequences $ (a_1, a_2, \ldots, a_r) \in \mathbb{F}_{q^m} $ where not all $ a_1, a_2, \ldots, a_r $ are equal, see \cite[Ex. 1.15.3]{leroy-noncommutative}. We have the following characterization, given in \cite[Cor. 45]{skew-multi}.

\begin{theorem} [\textbf{Characterization} \cite{skew-multi}] \label{th_char_multiplicity}
Let $ \mathbf{a} = (a_1, a_2, $ $ \ldots, a_r) \in \mathbb{F}_{q^m}^r $, where $ r \geq 2 $. Then $ \mathbf{a} $ is a multiplicity sequence if, and only if, either $ a_1 = a_2 = \ldots = a_r = 0 $ or $ a_1 \neq 0 $ and, for $ i \in [r-1] $, there exists $ \beta_i \in \mathbb{F}_{q^m}^* $ such that
$$ a_{i+1} = a_i^{\beta_i} \quad \textrm{and} \quad {\rm Tr}_q \left( \beta_i a_i^{-1} \right) \neq 0, $$
where $ {\rm Tr}_q $ denotes the trace of $ \mathbb{F}_{q^m} $ over $ \mathbb{F}_q $.
\end{theorem}

In particular, for any $ a_1 \in \mathbb{F}_{q^m} $ and any $ r \in \mathbb{Z}_+ $, there exists a multiplicity sequence $ (a_1, a_2, \ldots, a_r) \in \mathbb{F}_{q^m}^r $. Hence we may consider any multiplicity order over any point in $ \mathbb{F}_{q^m} $.

A trivial but important observation is that multiplicity sequences are monotonous.

\begin{proposition} [\textbf{Monotonicity}]
If $ (a_1, a_2, \ldots, a_r) \in \mathbb{F}_{q^m}^r $ is a multiplicity sequence, then so are the sequences $ (a_1, a_2, \ldots, a_i) \in \mathbb{F}_{q^m}^i $, for $ i \in [r-1] $.
\end{proposition}

Using multiplicity sequences, we may define arithmetic Hasse derivatives and multiplicities of skew polynomials as follows. The following is \cite[Def. 15]{skew-multi} and is consistent by applying right Euclidean division recursively.

\begin{definition} [\textbf{Arithmetic Hasse derivative} \cite{skew-multi}] \label{def_hasse_arithmetic}
Let $ r \in \mathbb{Z}_+ $ and let $ \mathbf{a}_r = (a_1, a_2, \ldots, a_r) $ $ \in \mathbb{F}_{q^m}^r $ be a multiplicity sequence. For every $ F \in \mathbb{F}_{q^m}[x;\sigma] $, there exist a unique $ G \in \mathbb{F}_{q^m}[x;\sigma] $ and unique coefficients $ D_{\mathbf{a}_i}(F) \in \mathbb{F}_{q^m} $, for $ i \in [r] $, such that
$$ F = G P_{\mathbf{a}_r} + D_{\mathbf{a}_{r}} (F) P_{\mathbf{a}_{r-1}} + \cdots + D_{\mathbf{a}_{2}} (F) P_{\mathbf{a}_1} + D_{\mathbf{a}_1} (F) , $$ 
where $ \mathbf{a}_i = (a_1, a_2, \ldots, a_i) \in \mathbb{F}_{q^m}^i $, for $ i \in [r] $. We call $ D_{\mathbf{a}_r}(F) $ the $ (r-1) $th Hasse derivative of $ F $ in $ a_1 $ (with respect to $ \mathbf{a}_r $). In particular, $ D_{\mathbf{a}_1}(F) = F(a_1) $ (the $ 0 $th derivative).
\end{definition}

By combining Definitions \ref{def_multiplicity} and \ref{def_hasse_arithmetic}, we deduce the following derivative criterion for multiplicity.

\begin{lemma} [\textbf{Arithmetic derivative criterion}] \label{lemma_derivative_criterion}
Given a multiplicity sequence $ \mathbf{a}_r = (a_1, a_2, $ $ \ldots, a_r) \in \mathbb{F}_{q^m}^r $, we have that $ a_1 $ is a zero of $ F $ of multiplicity $ r $ (via $ \aaa_r $) if and only if 
$$ D_{\mathbf{a}_1} (F) = D_{\mathbf{a}_2} (F) = \ldots = D_{\mathbf{a}_r} (F) = 0 , $$
where $ \mathbf{a}_i = (a_1, a_2, \ldots, a_i) \in \mathbb{F}_{q^m}^i $, for $ i \in [r] $.
\end{lemma}

With this type of multiplicities, we may upper bound the number of zeros with multiplicity by the degree. The following result is \cite[Th. 7]{skew-multi}.

\begin{theorem} [\textbf{Multiplicity-degree bound} \cite{skew-multi}] \label{th_multiplicity_bound}
Let $ a_1, a_2 , \ldots, a_n \in \mathbb{F}_{q^m} $ be distinct and let $ r_1,r_2, \ldots, r_n \in \mathbb{Z}_+ $. Then $ \Omega = \{ a_1, a_2, \ldots, a_n \} $ is P-independent if and only if, for every nonzero $ F \in \mathbb{F}_{q^m}[x;\sigma] $ that has $ a_i $ as a root of multiplicity $ r_i $, for $ i \in [n] $, it holds that
$$ \sum_{i=1}^n r_i \leq \deg(F). $$ 
\end{theorem}

Notice however that Hasse derivatives as in Definition \ref{def_hasse_arithmetic} recover arithmetic evaluations, as in Definition \ref{def_ev_arithmetic}, but not evaluations that yield $ \mathbb{F}_q $-linear maps as in Definition \ref{def_ev_linearized}. We can obtain $ \mathbb{F}_q $-linear maps from Hasse derivatives by slightly modifying them, as follows. This definition is new to the best of our knowledge.

\begin{definition} [\textbf{Linearized Hasse derivatives}] \label{def_hasse_linear}
Given $ F = \sum_{i=0}^d f_i x^i \in \mathbb{F}_{q^m}[x;\sigma] $ and a multiplicity sequence $ \mathbf{a} = (a_1, a_2, \ldots, a_r) \in \mathbb{F}_{q^m}^r $, we define, for $ \beta \in \mathbb{F}_{q^m} $,
$$ F_\mathbf{a}(\beta) = \sum_{i=0}^d f_i D_\mathbf{a}(x^i) \sigma^i(\beta) . $$
\end{definition}

Clearly the map $ F_\mathbf{a} : \mathbb{F}_{q^m} \longrightarrow \mathbb{F}_{q^m} $ is $ \mathbb{F}_q $-linear. Moreover, for $ r = 1 $, the map $ F_\mathbf{a} $ recovers that in Definition \ref{def_ev_linearized} since $ D_{a_1}(x^i) = N_i(a_1) $.

In order to connect arithmetic and linearized Hasse derivatives (Definitions \ref{def_hasse_arithmetic} and \ref{def_hasse_linear}, respectively), we extend the concept of conjugacy to multiplicity sequences.

\begin{definition} [\textbf{Multiplicity and conjugacy}]
Given $ \mathbf{a} = (a_1, a_2, \ldots, a_r) \in \mathbb{F}_{q^m}^r $ and $ \beta \in \mathbb{F}_{q^m}^* $, we define
$$ \mathbf{a}^\beta = (a_1^\beta, a_2^{\sigma(\beta)}, \ldots, a_r^{\sigma^{r-1}(\beta)}) \in \mathbb{F}_{q^m}^r. $$
\end{definition}

We will next prove that the conjugate of a multiplicity sequence is again a multiplicity sequence. We need the following lemma.

\begin{lemma} \label{lemma P_a beta}
Given $ \mathbf{a} = (a_1, a_2, \ldots, a_r) \in \mathbb{F}_{q^m}^r $ and $ \beta \in \mathbb{F}_{q^m}^* $,
$$ P_{\mathbf{a}^\beta} \beta = \sigma^r(\beta) P_\mathbf{a}. $$
\end{lemma}
\begin{proof}
It follows by repeatedly applying the identity $ (x-a^\beta)\beta = \sigma(\beta)(x-a) $, for all $ a,\beta \in \mathbb{F}_{q^m} $ with $ \beta \neq 0 $.
\end{proof}

\begin{proposition}
Let $ \mathbf{a} \in \mathbb{F}_{q^m}^r $ and $ \beta \in \mathbb{F}_{q^m}^* $. Then $ \mathbf{a} $ is a multiplicity sequence if and only if $ \mathbf{a}^\beta $ is a multiplicity sequence.
\end{proposition}
\begin{proof}
We only prove the direct implication, since the converse is obtained by applying $ \beta^{-1} $. Assume that $ \mathbf{a} $ is a multiplicity sequence and let $ b \in \mathbb{F}_{q^m} $ be such that $ P_{\mathbf{a}^\beta}(b) = 0 $. Thus by Lemma \ref{lemma P_a beta} and the product rule (Lemma \ref{lemma_product_rule}), 
$$ \sigma^r(\beta) P_\mathbf{a}(b^{\beta^{-1}}) = (P_{\mathbf{a}^\beta}\beta)(b^{\beta^{-1}}) = P_{\mathbf{a}^\beta}(b)\beta = 0. $$
Hence we deduce that $ P_{\mathbf{a}}(b^{\beta^{-1}}) = 0 $. Since $ \mathbf{a} $ is a multiplicity sequence, then $ b^{\beta^{-1}} = a_1 $, which is equivalent to $ b = a_1^\beta $. Thus the only zero of $ P_{\mathbf{a}^\beta} $ is $ a_1^\beta $, and we are done.
\end{proof}

We may now show the connection between both types of Hasse derivatives, which recovers Proposition \ref{prop_connection_evals} when $ r = 1 $.

\begin{proposition} [\textbf{Connecting derivatives}] \label{prop_connection_hasse_derivatives}
Given $ F \in \mathbb{F}_{q^m}[x;\sigma] $, a multiplicity sequence $ \mathbf{a} = (a_1, a_2, \ldots, a_r) \in \mathbb{F}_{q^m}^r $ and $ \beta \in \mathbb{F}_{q^m}^* $, we have that
$$ F_\mathbf{a}(\beta) = D_{\mathbf{a}^\beta}(F) \sigma^{r-1}(\beta). $$
In particular, $ F_\mathbf{a}(\beta) = 0 $ if and only if $ D_{\mathbf{a}^\beta}(F) = 0 $.
\end{proposition}
\begin{proof}
Since both types of Hasse derivatives are linear on skew polynomials, we only need to prove the formula for $ F = x^i $, for $ i \in \mathbb{N} $. Denoting $ \mathbf{a}_j = (a_1, a_2, \ldots, a_j) \in \mathbb{F}_{q^m}^j $, for $ j \in [r] $, we deduce from Lemma \ref{lemma P_a beta} that
\begin{equation*}
\begin{split}
x^i \beta & = \sum_{j=1}^r D_{\mathbf{a}_j^\beta}(x^i) P_{\mathbf{a}_{j-1}^\beta} \beta + G P_{\mathbf{a}^\beta} \beta \\
 & = \sum_{j=1}^r D_{\mathbf{a}_j^\beta}(x^i) \sigma^{j-1}(\beta) P_{\mathbf{a}_{j-1}} + G \sigma^r(\beta) P_\mathbf{a},
\end{split}
\end{equation*}
for some $ G \in \mathbb{F}_{q^m}[x;\sigma] $. Now, we also have 
$$ \sigma^i(\beta)x^i = \sum_{j=1}^r \sigma^i(\beta) D_{\mathbf{a}_j}(x^i) P_{\mathbf{a}_{j-1}} + H P_\mathbf{a}, $$
for some $ H \in \mathbb{F}_{q^m}[x;\sigma] $. Since $ x^i \beta = \sigma^i(\beta)x^i $, then by the uniqueness of quotients and remainders in the right Euclidean division, we deduce for $ j = r $ that
$$ \sigma^i(\beta) D_{\mathbf{a}_r}(x^i) = D_{\mathbf{a}_r^\beta}(x^i) \sigma^{r-1}(\beta) . $$
Since the left-hand side is $ F_\mathbf{a}(\beta) $, for $ F = x^i $, and the right-hand side is $ D_{\mathbf{a}^\beta}(F) \sigma^{r-1}(\beta) $, for $ F = x^i $, the result follows.
\end{proof}

This means that we may use the $ \mathbb{F}_q $-linear maps $ F_\mathbf{a} $ to count multiplicities of P-independent zeros. This will be crucial when working with linearized multiplicity codes.

Thanks to these definitions and results, we can give the following common generalization of linearized Reed--Solomon codes \cite{linearizedRS} and univariate multiplicity codes \cite{nrt}.

\begin{definition} [\textbf{Linearized multiplicity codes}] \label{def_lin_mult_codes}
Let $ a_1, a_2, \ldots, a_\ell \in \mathbb{F}_{q^m}^* $ be pairwise nonconjugate. Let also $ \beta_{i,1}, \beta_{i,2}, \ldots, \beta_{i,n_i} \in \mathbb{F}_{q^m} $ be $ \mathbb{F}_q $-linearly independent, for $ i \in [\ell] $. Let $ r \in \mathbb{Z}_+ $ and take nested multiplicity sequences $ \mathbf{a}_{i,j} = (a_{i,1}, a_{i,2}, \ldots, a_{i,j}) \in \mathbb{F}_{q^m}^j $ with $ a_{i,1} = a_i $, for $ j \in [r] $ and $ i \in [\ell] $. For $ F \in \mathbb{F}_{q^m}[x;\sigma] $, we define
$$
C(F) = \left(
\begin{array}{ccc|c|ccc}
F_{\mathbf{a}_{1,1}}(\beta_{1,1})
& \ldots
& F_{\mathbf{a}_{1,1}}(\beta_{1,n_1})
& \ldots
& F_{\mathbf{a}_{\ell,1}}(\beta_{\ell,1})
& \ldots
& F_{\mathbf{a}_{\ell,1}}(\beta_{\ell,n_\ell})
\\
F_{\mathbf{a}_{1,2}}(\beta_{1,1})
& \ldots
& F_{\mathbf{a}_{1,2}}(\beta_{1,n_1})
& \ldots
& F_{\mathbf{a}_{\ell,2}}(\beta_{\ell,1})
& \ldots
& F_{\mathbf{a}_{\ell,2}}(\beta_{\ell,n_\ell})
\\
\vdots
& \ddots
& \vdots
& \ddots
& \vdots
& \ddots
& \vdots
\\
F_{\mathbf{a}_{1,r}}(\beta_{1,1})
& \ldots
& F_{\mathbf{a}_{1,r}}(\beta_{1,n_1})
& \ldots
& F_{\mathbf{a}_{\ell,r}}(\beta_{\ell,1})
& \ldots
& F_{\mathbf{a}_{\ell,r}}(\beta_{\ell,n_\ell})
\end{array}
\right)
\in \mathbb{F}_{q^m}^{r \times n},
$$
where $ n = n_1 + n_2 + \cdots + n_\ell $. For $ k \in [rn] $, we define the linearized multiplicity code
$$ \M_k(\mathbf{a},\boldsymbol\beta) = \{ C(F) : F \in \mathbb{F}_{q^m}[x;\sigma], \deg(F) < k \} \subseteq \F_{q^m}^{r \times n} , $$
which is an $ \mathbb{F}_{q^m} $-linear code. The complete evaluation data is 
$$ \aaa = (\aaa_{1,r}, \aaa_{2,r}, \ldots, \aaa_{\ell,r}) \quad \textrm{and} \quad \boldsymbol\beta = (\beta_{1,1}, \ldots, \beta_{1,n_1}, \ldots, \beta_{\ell,1}, \ldots, \beta_{\ell, n_\ell}). $$
\end{definition}

Observe that, since $ a_1, a_2, \ldots, a_\ell \in \mathbb{F}_{q^m}^* $ must be pairwise nonconjugate and $ \beta_{i,1}, \beta_{i,2}, \ldots, $ $ \beta_{i,n_i} $ $ \in \mathbb{F}_{q^m} $ must be $ \mathbb{F}_q $-linearly independent, for $ i \in [\ell] $, a linearized multiplicity code has the restrictions
\begin{equation}
\ell \leq q-1 \quad \textrm{and} \quad n_i \leq m, \quad \forall i \in [\ell],
\label{eq_lin_mult_code_restrictions}
\end{equation}
where equalities can be attained. That is, we may choose $ \ell = q-1 $ and $ m = n_1 = n_2 = \ldots = n_\ell $ (notice that $ r $ is not involved in the restrictions (\ref{eq_lin_mult_code_restrictions})).

If $ r = 1 $ and $ \ell, m, n_1,n_2, \ldots, n_\ell \in \Z_+ $ are arbitrary (satisfying (\ref{eq_lin_mult_code_restrictions})), then linearized multiplicity codes become linearized Reed--Solomon codes \cite[Def. 31]{linearizedRS}. 

If $ m=n_1=n_2= \ldots = n_\ell = 1 $ and $ \ell , r \in \Z_+ $ is arbitrary (satisfying (\ref{eq_lin_mult_code_restrictions})), then linearized multiplicity codes become classical univariate multiplicity codes \cite{nrt} (see also \cite[Def. 3.4]{multiplicity}). This is because, in this case, $ \mathbb{F}_{q^m}[x;\sigma] = \mathbb{F}_q[x] $ and linearized Hasse derivatives and multiplicities of zeros of skew polynomials become, respectively, classical Hasse derivatives \cite{hasse} and classical multiplicities of zeros of polynomials. Notice however that in this case, we may add the zero and infinity evaluation points (i.e., $ \ell = q+1 $ may be attained).

If $ r=m=n_1=n_2= \ldots = n_\ell = 1 $ (the intersection of the above two cases), then linearized multiplicity codes become classical Reed--Solomon codes \cite{reed-solomon}.

\begin{remark} \label{remark_lin_mult_gabidulin}
For $ \ell=1 $, $ a_1 = 1 \in \mathbb{F}_{q^m} $ and arbitrary $ r,m,n \in \Z_+ $ (with $ n \leq m $), linearized multiplicity codes become a generalization of Gabidulin codes \cite{gabidulin} that consists on considering linearized Hasse derivatives of classical linearized polynomials. This generalization of Gabidulin codes is new to the best of our knowledge. 
\end{remark}

\section{The ordered sum-rank metric and MOSRD codes} \label{sec_ordered_SR_metric}

In this section, we introduce the ordered sum-rank metric (both orientations), provide its Singleton bound, define maximum ordered sum-rank distance (MOSRD) codes and prove that linearized multiplicity codes are MOSRD (and hence also ordered MDS). 

For the remainder of the manuscript, we consider positive integers $ r,m,\ell,n_1,n_2, \ldots, n_\ell $ and $ n = n_1 +n_2+ \cdots + n_\ell $, and consider the ambient vector space (over $ \F_{q^m} $)
$$ \mathbb{F}_{q^m}^{r \times n} \cong \prod_{i=1}^\ell \mathbb{F}_{q^m}^{r \times n_i}. $$
For $ t,s \in \Z_+ $ and a matrix $ Y \in \F_{q^m}^{t \times s} $, we denote by $ \col(Y) \subseteq \F_{q^m}^t $ the $ \F_q $-linear subspace generated by the $ s $ columns of $ Y $, and we define
\begin{equation}
\wt_R(Y) = \dim_{\F_q}(\col(Y)).
\label{eq_def_wtR}
\end{equation}
Notice that the columns of $ Y $ lie in the $ \F_q $-vector space $ \F_{q^m}^t \cong \F_q^{mt} $, and therefore the above weight is the rank of $ Y $ seen as a matrix in $ \F_q^{(mt) \times s} $ (see also Definition \ref{def_matrix_representation}).

\begin{definition} [\textbf{Ordered sum-rank metrics}] \label{def_OSR}
Whenever we write $ X = (X_1, X_2, \ldots, X_\ell) \in \F_{q^m}^{r \times n} $, we implicitly assume $ X_i \in \F_{q^m}^{r \times n_i} $, for $ i \in [\ell] $. We define the positively and negatively ordered sum-rank weights as $ \wt_{SR}^+, \wt_{SR}^- : \F_{q^m}^{r \times n} \longrightarrow \N $, respectively, given by
$$ \wt_{SR}^+ (X) = \sum_{i=1}^\ell \sum_{j=1}^r \wt_R(X_i^{[j]}) \quad \textrm{and} \quad \wt_{SR}^-(X) = \sum_{i=1}^\ell \sum_{j=1}^r \wt_R(X_i^{(j)}) , $$
where $ X = (X_1,X_2, \ldots, X_\ell) \in \F_{q^m}^{r \times n} $, and for $ j \in [r] $ and $ i \in [\ell] $,
$$ X_i = \left( \begin{array}{c}
\xx_{i,1} \\
\xx_{i,2} \\
\vdots \\
\xx_{i,r}
\end{array} \right) \in \mathbb{F}_{q^m}^{r \times n_i}, \quad X_i^{[j]} = \left( \begin{array}{c}
\xx_{i,1} \\
\xx_{i,2} \\
\vdots \\
\xx_{i,j}
\end{array} \right) \in \mathbb{F}_{q^m}^{j \times n_i} \quad \textrm{and} \quad X_i^{(j)} = \left( \begin{array}{c}
\xx_{i,j} \\
\xx_{i,j+1} \\
\vdots \\
\xx_{i,r}
\end{array} \right) \in \mathbb{F}_{q^m}^{(r-j+1) \times n_i}. $$
The associated metrics are, respectively, $ \dd_{SR}^+, \dd_{SR}^- : (\F_{q^m}^{r \times n})^2 \longrightarrow \N $, given for $ X,Y \in \F_{q^m}^{r \times n} $ by 
$$ \dd_{SR}^+(X,Y) = \wt_{SR}^+(X-Y) \quad \textrm{and} \quad \dd_{SR}^-(X,Y) = \wt_{SR}^-(X-Y) . $$
As usual, for a code (i.e., a subset) $ \C \subseteq \F_{q^m}^{r \times n} $ and a metric $ \dd $, we define the minimum distance $ \dd(\C) = \min \{ \dd(X,Y) : X,Y \in \C, X \neq Y \} $ if $ |\C| > 1 $. If $ \C $ is linear (i.e., $ \F_{q^m} $-linear) and $ \dd $ is given by a weight $ \wt $ as above, then $ \dd(\C) = \min \{ \wt(X) : X \in \C \setminus \{ 0 \} \} $.
\end{definition}
%

It is straightforward to prove that both sum-rank metrics are indeed metrics. From now on, we will refer to the positively ordered sum-rank metric simply as the ordered sum-rank metric. 
%
%

Clearly, if $ r=1 $, then both ordered sum-rank metrics coincide and give the classical sum-rank metric, introduced in \cite[Sec. III-D]{multishot} and implicitly considered earlier in \cite[Sec. III]{space-time-kumar}. In fact, for $ X = (X_1,X_2, \ldots, X_\ell) \in \F_{q^m}^{r \times n} $, we have
\begin{equation}
\wt_{SR}^+(X) = \sum_{j=1}^r \sum_{i=1}^\ell \wt_R(X_i^{[j]}) = \sum_{j=1}^r \wt_{SR}(X^{[j]}),
\label{eq_SR_weight_ordered_SR}
\end{equation}
which in particular implies that $ \wt_{SR}^+(X) \in [0,rn] $ since $ \wt_{SR}(X^{[j]}) \in [0,n] $, for $ j \in [r] $. For this reason, we may also use the convention $ \dd_{SR}^+(\C) = rn+1 $ and $ \dd_{SR}(\C) = n + 1 $ if $ |\C| = 1 $.

Now, if $ m=n_1=n_2= \ldots = n_\ell = 1 $, then the ordered sum-rank metric recovers the Niederreiter-Rosenbloom-Tsfasman (NRT) metric, introduced in \cite{niederreiter} and later in \cite[Sec. 1]{nrt}. To see this, notice that in that case, $ X = (X_1,X_2, \ldots, X_\ell) \in \F_q^{r \times \ell} $, 
$$ X_i = \left( \begin{array}{c}
x_{i,1} \\
x_{i,2} \\
\vdots \\
x_{i,r}
\end{array} \right) \in \F_q^{r \times 1} \quad \textrm{and} \quad \wt_R(X_i^{[j]}) = \left\lbrace \begin{array}{ll}
0 & \textrm{ if } j < t_i, \\
1 & \textrm{ if } j \geq t_i,
\end{array} \right. $$
where $ t_i = \min \{ u \in [r] : x_{i,u} \neq 0 \} $ (defined as $ t_i = r+1 $ if $ X_i = 0 $), for $ i \in [\ell] $. Hence, in this case,
$$ \wt_{SR}^+ (X) = \sum_{i=1}^\ell (r-t_i+1), $$
which is the definition of the NRT weight. The negatively ordered sum-rank metric coincides with the NRT metric but for the opposite order of column indices \cite{poset}. The NRT metric is also known as the ordered Hamming metric \cite{barg_poset} since it can also be expressed as in (\ref{eq_SR_weight_ordered_SR}). Let $ X = (X_1,X_2, \ldots, X_\ell) \in \F_q^{r \times \ell} $ and define $ \wt_H(X^{[j]}) = |\{ i \in [\ell] : X_i^{[j]} \neq 0 \} | $ (i.e., the Hamming weight of $ X^{[j]} \in \F_q^{j \times \ell} $ with alphabet $ \F_q^j $ and length $ \ell $), for $ j \in [r] $. Defining $ t_i $ as above, for $ i \in [\ell] $, the NRT weight of $ X $ satisfies
\begin{equation}
\wt_{NRT}(X) = \sum_{i=1}^\ell (r-t_i+1) = \sum_{j=1}^r \wt_H(X^{[j]}).
\label{eq_NRT_weight_ordered_Hamming}
\end{equation}
For the NRT metric with the opposite order, we would simply replace $ X^{[j]} $ by $ X^{(j)} $.

\begin{remark} [\textbf{The ordered rank metric}] \label{remark_ordered_rank}
For $ \ell = 1 $, the ordered sum-rank metrics particularize to a definition of ordered rank metrics that is new to the best of our knowledge.  
\end{remark}

\begin{remark} 
As shown in Figure \ref{fig_metrics}, the intersection between the ordered sum-rank metric and subspace-generated metrics \cite{firer_poset, gabidulin_metrics} is exactly the sum-rank metric. This is because with a subspace-generated metric, the ambient space is generated by vectors of weight $ 1 $, whereas this is possible for the ordered sum-rank metric if and only if $ r = 1 $ (the sum-rank case). It is also not hard to prove that the intersection between the ordered sum-rank metric and the poset metrics \cite{poset, firer_poset} is the NRT metric. 
\end{remark}

We now give a very useful characterization of the ordered sum-rank metric as minimums of NRT metrics after blockwise multiplication of invertible matrices. When $ r = 1 $, the next result becomes \cite[Th. 1.2]{fnt} (first obtained in the proof of \cite[Th. 3]{linearizedRS}).

\begin{theorem} \label{th_OSR_min}
For all $ X = (X_1, X_2, \ldots, X_\ell) \in \F_{q^m}^{r \times n} $, we have that
$$ \wt_{SR}^+(X) = \min \{ \wt_{NRT}(X_1A_1,X_2A_2, \ldots, X_\ell A_\ell) : A_i \in \GL_{n_i}(\F_q), i \in [\ell] \}, $$
where $ \wt_{NRT} $ denotes the NRT weight in $ \F_{q^m}^{r \times n} $, given as in (\ref{eq_NRT_weight_ordered_Hamming}) by 
$$ \wt_{NRT}(X) = \sum_{j=1}^r \wt_H(X^{[j]}), $$
where $ \wt_H(X^{[j]}) $ is the number of nonzero columns of $ X^{[j]} \in \F_{q^m}^{j \times n} $, for $ j \in [r] $. In particular, for a (linear or nonlinear) code $ \C \subseteq \F_{q^m}^{r \times n} $,
$$ \dd_{RS}^+(\C) = \min \{ \dd_{NRT}(\C \diag(A_1,A_2, \ldots, A_\ell)) : A_i \in \GL_{n_i}(\F_q), i \in [\ell] \}. $$
The same characterizations hold for $ \wt_{SR}^- $ and $ \dd_{SR}^- $ (using the reversed order for the NRT metric).
\end{theorem}
\begin{proof}
Since the rank of a matrix is upper bounded by the number of its nonzero columns, then
$$ \wt_{SR}^+(X) = \sum_{i=1}^\ell \sum_{j=1}^r \wt_R(X_i^{[j]}) \leq \sum_{i=1}^\ell \sum_{j=1}^r \wt_H(X_i^{[j]}) = \wt_{NRT}(X). $$
For $ A_i \in \GL_{n_i}(\F_q) $, since multiplying on the right by $ A_i $ does not change ranks, for $ i \in [\ell] $, we deduce that
$$ \wt_{SR}^+(X) = \wt_{SR}^+ (X_1A_1, X_2A_2, \ldots, X_\ell A_\ell) \leq \wt_{NRT}(X_1A_1, X_2A_2, \ldots, X_\ell A_\ell). $$
We next show that equality is achieved for certain invertible matrices. Given $ i \in [\ell] $ and $ j \in [r] $, let $ \varphi_{i,j} : \F_q^{n_i} \longrightarrow \F_{q^m}^{j \times 1} $ be the $ \F_q $-linear map given by $ \varphi_{i,j}(\aaa) = X_i^{[j]} \aaa^\intercal $, for $ \aaa \in \F_q^{n_i} $. Define $ \K_{i,j} = \ker(\varphi_{i,j}) \subseteq \F_q^{n_i} $ and $ k_{i,j} = \dim_{\F_q}(\K_{i,j}) $. Since we have the flag of subspaces
$$ \K_{i,r} \subseteq \K_{i,r-1} \subseteq \ldots \subseteq \K_{i,2} \subseteq \K_{i,1} \subseteq \F_q^{n_i}, $$
we may choose a basis $ \aaa_{i,1},\aaa_{i,2}, \ldots, \aaa_{i,n_i} $ of $ \F_q^{n_i} $ such that the first $ k_{i,j} $ vectors form a basis of $ \K_{i,j} $. Let $ A_i \in \GL_{n_i}(\F_q) $ be the matrix with columns $ \aaa_{i,1}^\intercal, \aaa_{i,2}^\intercal, \ldots, \aaa_{i,n_i}^\intercal $. Then by construction, exactly the first $ k_{i,j} $ columns of $ X_i^{[j]} A_i $ are zero. Therefore
$$ \wt_H(X_i^{[j]}A_i) = n_i - k_{i,j} = \wt_R(X_i^{[j]}), $$
for $ i \in [\ell] $ and $ j \in [r] $. We conclude that
$$ \wt_{NRT}(X_1A_1, X_2A_2, \ldots, X_\ell A_\ell) = \sum_{i=1}^\ell \sum_{j=1}^r \wt_H(X_i^{[j]}A_i) = \sum_{i=1}^\ell \sum_{j=1}^r \wt_R(X_i^{[j]}) = \wt_{SR}^+(X). $$
\end{proof}

In particular, choosing identity matrices, we deduce that
\begin{equation}
\wt_{SR}^+(X) \leq \wt_{NRT}(X),
\label{eq_OSR_leq_NRT}
\end{equation}
for $ X \in \F_{q^m}^{r \times n} $. Therefore, we may deduce a Singleton bound for the ordered sum-rank metric from that for the NRT metric \cite[Th. 1]{nrt}.

\begin{corollary} [\textbf{Singleton bound}] \label{cor_singleton}
For a (linear or nonlinear) code $ \C \subseteq \F_{q^m}^{r \times n} $, we have that
\begin{equation}
\dd_{SR}^+(\C) \leq rn - k + 1, 
\label{eq_singleton}
\end{equation}
where $ k = \log_{q^m} | \C | $. Furthermore, equality holds if and only if, for all $ A_i \in \GL_{n_i}(\F_q) $, for $ i \in [\ell] $, $ \C \diag(A_1,A_2, \ldots, A_\ell) \subseteq \F_{q^m}^{r \times n} $ achieves the same Singleton bound for the NRT metric as defined in Theorem \ref{th_OSR_min}. Analogously for $ \dd_{SR}^- $.
\end{corollary}

\begin{definition} [\textbf{MOSRD codes}] \label{def_mosrd}
We say that a (linear or nonlinear) code $ \C \subseteq \F_{q^m}^{r \times n} $ is (positively) maximum ordered sum-rank distance (MOSRD) if it achieves equality in (\ref{eq_singleton}). Analogously for negatively MOSRD codes and $ \dd_{SR}^- $.
\end{definition}

In the literature, codes that attain the Singleton bound for the NRT metric are sometimes called (positively) ordered MDS \cite{barg_park} (analogously for the negative orientation). Thus we deduce the following from Corollary \ref{cor_singleton}.

\begin{corollary} \label{cor_MOSRD_iff_ordered_MDS}
Fixing an orientation, a (linear or nonlinear) code $ \C \subseteq \F_{q^m}^{r \times n} $ is MOSRD if and only if $ \C \diag(A_1,A_2, $ $ \ldots, $ $ A_\ell) \subseteq \F_{q^m}^{r \times n} $ is ordered MDS, for all $ A_i \in \GL_{n_i}(\F_q) $, for $ i \in [\ell] $. In particular, in that case, $ \C $ itself is ordered MDS.
\end{corollary}

We may now easily deduce that linearized multiplicity codes are MOSRD (and thus ordered MDS by Corollary \ref{cor_MOSRD_iff_ordered_MDS}).

\begin{theorem} \label{th_lin_mult_mosrd}
With notation and assumptions as in Definition \ref{def_lin_mult_codes}, the linearized multiplicity code $ \M_k(\mathbf{a},\boldsymbol\beta) \subseteq \F_{q^m}^{r \times n} $ is a linear (positively) MOSRD code of dimension $ k $ (over $ \F_{q^m} $), for all $ k \in [rn] $. In particular, they are (positively) ordered MDS.
\end{theorem}
\begin{proof}
Let $ A_i \in \GL_{n_i}(\F_q) $, for $ i \in [\ell] $. Since linearized Hasse derivatives are clearly $ \F_q $-linear (Definition \ref{def_hasse_linear}), it is easy to see that
$$ \M_k(\aaa,\boldsymbol\beta) \diag(A_1,A_2, \ldots, A_\ell) = \M_k(\aaa, \boldsymbol\beta \diag(A_1,A_2, \ldots, A_\ell)). $$
Since the $ n_i $ components of $ (\beta_{i,1},\beta_{i,2},\ldots, \beta_{i,n_i}) A_i \in \F_{q^m}^{n_i} $ are again $ \F_q $-linearly independent, for $ i \in [\ell] $, then by Corollary \ref{cor_singleton}, we only need to prove that $ \M_k(\aaa,\boldsymbol\beta) $ has dimension $ k $ and is ordered MDS.

Let $ F \in \F_{q^m}[x;\sigma] $ be nonzero with $ \deg(F) < k $. For all $ h \in [n_i] $ and $ i \in [\ell] $, let
$$ t_{i,h} = \max \{ t \in [r] : F_{\aaa_{i,j}}(\beta_{i,h}) = 0 , \forall j \in [t] \}, $$
and define $ t_{i,h} = 0 $ if $ F_{\aaa_{i,1}}(\beta_{i,h}) \neq 0 $. From the definition of the NRT metric in $ \F_{q^m}^{r \times n} $ (see Theorem \ref{th_OSR_min} and (\ref{eq_NRT_weight_ordered_Hamming})), we deduce that
$$ \wt_{NRT}(C(F)) = \sum_{j=1}^r \wt_H(C(F)^{[j]}) = \sum_{i=1}^\ell \sum_{h=1}^{n_i} (r-t_{i,h}) = rn - \sum_{i=1}^\ell \sum_{h=1}^{n_i} t_{i,h}. $$
By Lemma \ref{lemma_derivative_criterion} and Proposition \ref{prop_connection_hasse_derivatives}, $ F $ has $ a_i^{\beta_{i,h}} $ as a zero of multiplicity $ t_{i,h} $, for $ h \in [n_i] $ and $ i \in [\ell] $. Thus by Theorem \ref{th_multiplicity_bound}, 
$$ \sum_{i=1}^\ell \sum_{h=1}^{n_i} t_{i,h} \leq \deg(F). $$
Hence we conclude that
$$ \wt_{NRT}(C(F)) = rn - \sum_{i=1}^\ell \sum_{h=1}^{n_i} t_{i,h} \geq rn - \deg(F) \geq rn - k + 1. $$
This shows that $ \M_k(\mathbf{a},\boldsymbol\beta) $ has dimension $ k $ and is ordered MDS (and thus MOSRD as discussed above). 
\end{proof}

\section{Supports, duality and MSRD consequences} \label{sec_supports}

In this section, we extend the theory of supports for the sum-rank metric \cite{gsrws} to the ordered sum-rank metric. We will treat both orientations since, as in the case of the NRT metric, the dual of a positively MOSRD code will be a negatively MOSRD code (and viceversa). We will conclude the section by showing that MOSRD codes are also dually quasi MSRD \cite{qmsrd} and, if their dimension is divisible by $ r $, then they are MSRD and hence MDS.

\begin{definition} [\textbf{Matrix representation}] \label{def_matrix_representation}
Fix an ordered basis $ \aalpha = (\alpha_1, \alpha_2, \ldots, \alpha_m) \in \F_{q^m}^m $ of $ \F_{q^m} $ over $ \F_q $. For each $ s \in \Z_+ $, define the matrix representation map $ M_{\aalpha} : \F_{q^m}^s \longrightarrow \F_q^{m \times s} $ by 
$$ M_{\aalpha} (c_1,c_2, \ldots, c_s) = \left( \begin{array}{cccc}
c_{1,1} & c_{1,2} & \ldots & c_{1,s} \\
c_{2,1} & c_{2,2} & \ldots & c_{2,s} \\
\vdots & \vdots & \ddots & \vdots \\
c_{m,1} & c_{m,2} & \ldots & c_{m,s} 
\end{array} \right) \in \F_q^{m \times s}, $$
for $ (c_1,c_2, \ldots, c_s) \in \F_{q^m}^s $, where $ c_{i,j} \in \F_q $ are the unique scalars such that $ c_j = \sum_{i=1}^m c_{i,j} \alpha_i $, for $ i \in [m] $ and $ j \in [s] $. For $ t,s \in \Z_+ $, we extend this map to $ M_{\aalpha} : \F_{q^m}^{t \times s} \longrightarrow \F_q^{(mt) \times s} $ by
$$ M_{\aalpha} \left( \begin{array}{cccc}
x_{1,1} & x_{1,2} & \ldots & x_{1,s} \\
x_{2,1} & x_{2,2} & \ldots & x_{2,s} \\
\vdots & \vdots & \ddots & \vdots \\
x_{t,1} & x_{t,2} & \ldots & x_{t,s} 
\end{array} \right) = \left( \begin{array}{cccc}
M_{\aalpha}(x_{1,1}) & M_{\aalpha}(x_{1,2}) & \ldots & M_{\aalpha}(x_{1,s}) \\
M_{\aalpha}(x_{2,1}) & M_{\aalpha}(x_{2,2}) & \ldots & M_{\aalpha}(x_{2,s}) \\
\vdots & \vdots & \ddots & \vdots \\
M_{\aalpha}(x_{t,1}) & M_{\aalpha}(x_{t,2}) & \ldots & M_{\aalpha}(x_{t,s}) 
\end{array} \right), $$
for all $ x_{i,j} \in \F_{q^m} $, for $ i \in [t] $ and $ j \in [s] $. 
\end{definition}

Essentially, the map $ M_{\aalpha} : \F_{q^m}^{t \times s} \longrightarrow \F_q^{(mt) \times s} $ expands vertically every entry in $ \F_{q^m} $ as a column vector in $ \F_q^m $. Clearly the second map becomes the first one when $ t=1 $. However, we use the same notation for simplicity. All of these maps are $ \F_q $-linear vector space isomorphisms.

We now define the lattices of supports that are appropriate for the ordered sum-rank metrics (both orientations). From now on, we encode the length partition as $ \nn = (n_1, n_2, \ldots , n_\ell) \in \Z_+^\ell $. 

\begin{definition} [\textbf{Support lattices}] \label{def_support_lattices}
Let $ \PP^+_{q,\nn,r} $ and $ \PP^-_{q,\nn,r} $ be the families of tuples $ \LL = (\LL_{i,j})_{i=1,j=1}^{\ell,r} $ and $ \K = (\K_{i,j})_{i=1,j=1}^{\ell,r} $, respectively, such that
$$ \LL_{i,1} \subseteq \LL_{i,2} \subseteq \ldots \subseteq \LL_{i,r} \subseteq \F_q^{n_i} \quad \textrm{and} \quad \F_q^{n_i} \supseteq \K_{i,1} \supseteq \K_{i,2} \supseteq \ldots \supseteq \K_{i,r}, $$
and they are all $ \F_q $-vector subspaces, for $ i \in [\ell] $. For $ \LL = (\LL_{i,j})_{i=1,j=1}^{\ell,r} \in \PP^+_{q, \nn,r} $ and $ \K = (\K_{i,j})_{i=1,j=1}^{\ell,r} \in \PP^-_{q,\nn,r} $, we define
$$ \rk(\LL) = \sum_{i=1}^\ell \sum_{j=1}^r \dim_{\F_q}(\LL_{i,j}) \quad \textrm{and} \quad \rk(\K) = \sum_{i=1}^\ell \sum_{j=1}^r \dim_{\F_q}(\K_{i,j}). $$
\end{definition}

When $ r=1 $, Definition \ref{def_support_lattices} recovers the support lattices for the sum-rank metric from \cite[Def. 2 \& 3]{gsrws}. For any $ r \geq 1 $, one can define componentwise inclusions and operations in $ \PP^+_{q,\nn,r} $ and $ \PP^-_{q,\nn,r} $ as in \cite[Def. 3]{gsrws} that turn these sets into posets and, more strongly, into lattices (we treat duality in Proposition \ref{prop_support_duality}). See \cite[Ch. 3]{stanley} for the definitions of poset and (ordered) lattice.

We now turn to support spaces (i.e., subspaces of tuples whose supports are contained in a given flag of subspaces).

\begin{definition} [\textbf{Support spaces}] \label{def_support_spaces}
For $ X = (X_1,X_2, \ldots, X_\ell) \in \F_{q^m}^{r \times n} $, we define its positive and negative supports, respectively, as
\begin{equation*}
\begin{split}
\supp^+(X) = (\supp^+_{i,j}(X))_{i=1,j=1}^{\ell,r}, \quad \textrm{where} \quad & \supp^+_{i,j}(X) = \row(M_{\aalpha}(X_i^{[j]})) \subseteq \F_q^{n_i}, \\
\supp^-(X) = (\supp^-_{i,j}(X))_{i=1,j=1}^{\ell,r}, \quad \textrm{where} \quad & \supp^-_{i,j}(X) = \row(M_{\aalpha}(X_i^{(j)})) \subseteq \F_q^{n_i},
\end{split}
\end{equation*}
for $ i \in [\ell] $ and $ j \in [r] $, where $ \row(Y) \subseteq \F_q^s $ denotes the $ \F_q $-linear subspace generated by the rows of a matrix $ Y \in \F_q^{t \times s} $, for $ t,s \in \Z_+ $. For $ \LL \in \PP^+_{q,\nn,r} $ and $ \K \in \PP^-_{q,\nn,r} $, we define their positive and negative support spaces, respectively, as
$$ \V_\LL^+ = \{ X \in \F_{q^m}^{r \times n} : \supp^+(X) \subseteq \LL \} \quad \textrm{and} \quad \V_\K^- = \{ X \in \F_{q^m}^{r \times n} : \supp^-(X) \subseteq \K \}, $$
where inclusions in $ \PP^+_{q,\nn,r} $ and $ \PP^-_{q,\nn,r} $ are defined componentwise as in \cite[Def. 3]{gsrws}.
\end{definition}

As usual, the definition of support does not depend on the chosen ordered basis $ \aalpha \in \F_{q^m}^m $ of $ \F_{q^m} $ over $ \F_q $.

When $ r=1 $, Definition \ref{def_support_spaces} recovers the supports and support spaces for the sum-rank metric from \cite[Def. 4 \& 5]{gsrws}. When $ m = n_1 = n_2 = \ldots = n_\ell = 1 $, supports as in Definition \ref{def_support_spaces} recover support ideals for the NRT metric seen as a poset metric \cite{poset}. We have the following properties. 

\begin{proposition} \label{prop_supports_basic_props}
Let $ X \in \F_{q^m}^{r \times n} $, $ \LL \in \PP^+_{q,\nn,r} $ and $ \K \in \PP^-_{q,\nn,r} $. 
\begin{enumerate}
\item
$ \wt_{SR}^+(X) = \rk(\supp^+(X)) $ and $ \wt_{SR}^-(X) = \rk(\supp^-(X)) $.
\item
$ \V_\LL^+ $ and $ \V_\K^- $ are $ \F_{q^m} $-linear subspaces of $ \F_{q^m}^{r \times n} $ such that
$$ \dim_{\F_{q^m}}(\V_\LL^+) = \rk(\LL) \quad \textrm{and} \quad \dim_{\F_{q^m}}(\V_\K^-) = \rk(\K) . $$
\end{enumerate}
\end{proposition}
\begin{proof}
Item 1 follows directly from the definitions. Now we prove Item 2. For $ i \in [\ell] $, since $ \LL_{i,1} \subseteq \LL_{i,2} \subseteq \ldots \subseteq \LL_{i,r} \subseteq \F_q^{n_i} $, we may choose a basis $ \bb_{i,1}, \bb_{i,2}, \ldots, \bb_{i,n_i} $ of $ \F_q^{n_i} $ such that the first $ k_{i,j} = \dim_{\F_q}(\LL_{i,j}) $ elements form a basis of $ \LL_{i,j} $, for $ j \in [r] $. Now, notice that $ X \in \V_\LL^+ $ if and only if $ \row(M_{\aalpha}(X_i^{[j]})) \subseteq \LL_{i,j} $, for $ i \in [\ell] $ and $ j \in [r] $. Moreover, for $ \xx_{i,j} \in \F_{q^m}^{n_i} $, we have that $ \xx_{i,j} = \aalpha M_{\aalpha}(\xx_{i,j}) $, for $ i \in [\ell] $ and $ j \in [r] $. Thus it is easy to see that a basis of $ \V_\LL^+ $ (as an $ \F_{q^m} $-linear subspace) is given by the matrices whose $ i $th block of columns has $ \bb_{i,h} \in \F_q^{n_i} $ in the $ j $th row, and zeros everywhere else, for all $ h \in [k_{i,j}] $, all $ j \in [r] $ and all $ i \in [\ell] $. Thus 
$$ \dim_{\F_{q^m}}(\V_\LL^+) = \sum_{i=1}^\ell \sum_{j=1}^r k_{i,j} = \sum_{i=1}^\ell \sum_{j=1}^r \dim_{\F_q}(\LL_{i,j}), $$
and we are done. Similarly for $ \V_\K^- $.
\end{proof}

We now turn to duality of supports and support spaces. In $ \F_q^{n_i} $, for $ i \in [\ell] $, we consider the usual $ \F_q $-bilinear inner product, and in $ \F_{q^m}^{r \times n} $, we consider the usual $ \F_{q^m} $-bilinear inner product. That is, if $ X = (x_{j,h})_{j=1,h=1}^{r,n}, Y = (y_{j,h})_{j=1,h=1}^{r,n} \in \F_{q^m}^{r \times n} $, then their usual inner product is
\begin{equation}
\langle X , Y \rangle = {\rm Tr}(XY^\intercal) = \sum_{j=1}^r \sum_{h=1}^n x_{j,h} y_{j,h} \in \F_{q^m},
\label{eq_def_inner_product}
\end{equation}
where $ {\rm Tr} $ denotes the trace of a matrix. For an $ \F_{q^m} $-linear code $ \C \subseteq \F_{q^m}^{r \times n} $, we define its dual as $ \C^\perp = \{ Y \in \F_{q^m}^{r \times n} : \langle X,Y \rangle = 0, \forall X \in \C \} $ $ \subseteq \F_{q^m}^{r \times n} $. The corresponding definitions in $ \F_q^{n_i} $ are the same but taking $ r=m=1 $ and $ n = n_i $.

\begin{proposition} [\textbf{Support duality}] \label{prop_support_duality}
Given $ \LL = (\LL_{i,j})_{i=1,j=1}^{\ell,r} \in \PP^+_{q,\nn,r} $ and $ \K = (\K_{i,j})_{i=1,j=1}^{\ell,r} \in \PP^-_{q,\nn,r} $, we have that
\begin{equation*}
\begin{split}
\LL^\perp & = (\LL_{i,j}^\perp)_{i=1,j=1}^{\ell,r} \in \PP^-_{q,\nn,r}, \quad \rk(\LL^\perp) = rn - \rk(\LL) \quad \textrm{and} \quad (\V_\LL^+)^\perp = \V_{\LL^\perp}^-, \\
\K^\perp & = (\K_{i,j}^\perp)_{i=1,j=1}^{\ell,r} \in \PP^+_{q,\nn,r}, \quad \rk(\K^\perp) = rn - \rk(\K) \quad \textrm{and} \quad (\V_\K^-)^\perp = \V_{\K^\perp}^+.
\end{split}
\end{equation*}
Furthermore, taking duals induces a bijection between $ \PP^+_{q,\nn,r} $ and $ \PP^-_{q,\nn,r} $.
\end{proposition}
\begin{proof}
Taking duals induces a bijection between $ \PP^+_{q,\nn,r} $ and $ \PP^-_{q,\nn,r} $ since
$$ \LL_{i,1} \subseteq \LL_{i,2} \subseteq \ldots \subseteq \LL_{i,r} \subseteq \F_q^{n_i} \quad \textrm{implies} \quad \F_q^{n_i} \supseteq \LL_{i,1}^\perp \supseteq \LL_{i,2}^\perp \supseteq \ldots \supseteq \LL_{i,r}^\perp . $$
Fix $ i \in [\ell] $. As in the proof of Proposition \ref{prop_supports_basic_props}, choose a basis $ \bb_{i,1}, \bb_{i,2}, \ldots, \bb_{i,n_i} $ of $ \F_q^{n_i} $ such that the first $ k_{i,j} = \dim_{\F_q}(\LL_{i,j}) $ elements form a basis of $ \LL_{i,j} $, for $ j \in [r] $. We may similarly choose a basis $ \cc_{i,1},\cc_{i,2} , \ldots, \cc_{i,n_i} $ of $ \F_q^{n_i} $ such that the first $ k^\perp_{i,j} = \dim_{\F_q}(\LL_{i,j}^\perp) $ elements form a basis of $ \LL_{i,j}^\perp $, where $ k^\perp_{i,j} = n_i - k_{i,j} $, for $ j \in [r] $. In particular, we see that 
$$ \rk(\LL^\perp) = \sum_{i=1}^\ell \sum_{j=1}^r k^\perp_{i,j} = rn - \sum_{i=1}^\ell \sum_{j=1}^r k_{i,j} = rn - \rk(\LL). $$
The formula for $ \rk(\K^\perp) $ is proven similarly. Now looking at the bases of $ \V_\LL^+ $ and $ \V_{\LL^\perp}^- $ given, respectively, by $ \bb_{i,1}, \bb_{i,2}, \ldots, \bb_{i,n_i} $ and $ \cc_{i,1}, \cc_{i,2} , \ldots , \cc_{i,n_i} $, as in the proof of Proposition \ref{prop_supports_basic_props}, one can easily verify that they are orthogonal, which implies that
$$ \V_{\LL^\perp}^- \subseteq ( \V_\LL^+)^\perp . $$
Since their dimensions coincide, they are the same subspace. Similarly for $ (\V_\K^-)^\perp = \V_{\K^\perp}^+ $.
\end{proof}

We are now ready to show that the dual of an $ \F_{q^m} $-linear MOSRD code is again MOSRD (reversing orientations). For $ r=1 $ and the rest of parameters being arbitrary, this result becomes the well-known duality theorem for MSRD codes \cite[Th. 5]{gsrws}. For $ m = n_1= n_2= \ldots =n_\ell = 1 $ and arbitrary $ r \in \Z_+ $, this result becomes the duality theorem for ordered MDS codes, obtained in \cite[Cor. 3.5]{martin} and \cite[Th. 4.1(i)]{skriganov}. 

\begin{theorem} [\textbf{MOSRD duality}] \label{th_mosrd_duality}
Let $ \C \subseteq \F_{q^m}^{r \times n} $ be an $ \F_{q^m} $-linear code. Then $ \C $ is positively MOSRD if and only if $ \C^\perp $ is negatively MOSRD (and viceversa).
\end{theorem}
\begin{proof}
Let $ k = \dim_{\F_{q^m}}(\C) $, thus $ \dim_{\F_{q^m}}(\C^\perp) = rn - k $, and let $ d = \dd_{SR}^+(\C) $ (thus $ d = rn - k +1 $). Let $ Y \in \C^\perp $ be such that $ \wt_{SR}^-(Y) \leq k $. Then there exists $ \K \in \PP^-_{q,\nn,r} $ such that $ \supp^-(Y) \subseteq \K $ and $ \rk(\K) = k $. To see this, if $ \wt_{SR}^-(Y) < k \leq rn $, then for some $ i \in [\ell] $, there exists $ j \in [r] $ such that $ \supp^-_{i,j}(Y) \subsetneq \F_q^{n_i} $ and $ j = 1 $ or $ \supp^-_{i,j-1}(Y) = \F_q^{n_i} $ (chains in $ \PP^-_{q,\nn,r} $ are descending), and we simply need to increase the dimension of such subspaces $ \supp^-_{i,j}(Y) $, as needed. 

Define $ \LL = \K^\perp \in \PP^+_{q,\nn,r} $. By Propositions \ref{prop_supports_basic_props} and \ref{prop_support_duality}, $ (\V_\K^-)^\perp = \V_\LL^+ $,
$$ \dim_{\F_{q^m}}(\V_\K^-) = \rk(\K) = k \quad \textrm{and} \quad \dim_{\F_{q^m}}(\V_\LL^+) = \rk(\LL) = rn - k. $$
Since $ d = rn-k+1 $, it must hold that $ \C \cap \V_\LL^+ = \{ 0 \} $. Since their dimensions add up to $ rn $, it must hold that $ \F_{q^m}^{r \times n} = \C \oplus \V_\LL^+ $. Taking duals, $ \C^\perp \cap \V_\K^- = \{ 0 \} $. Hence $ Y = 0 $, and we conclude that $ \dd_{SR}^-(\C^\perp) \geq k+1 $, and $ \C^\perp $ is negatively MOSRD.
\end{proof}

We conclude the section showing that MOSRD (any orientation) implies dually quasi MSRD \cite{qmsrd} and, if the dimension is divisible by $ r $, then MOSRD implies MSRD \cite{linearizedRS} (and thus MDS). 

We recall the definition of quasi MSRD codes from \cite{qmsrd}. Let $ \C \subseteq \prod_{i=1}^\ell \F_q^{t \times n_i} \cong \F_q^{t \times n} $, where $ n = n_1 + n_2 + \cdots + n_\ell $, be an $ \F_q $-linear code with $ \kappa = \dim_{\F_q}(\C) $. If its minimum sum-rank distance (always according to the partition $ n = n_1 + n_2 + \cdots + n_\ell $) is $ d = \dd_{SR}(\C) $, then the Singleton bound says that $ \kappa \leq t(n - d + 1) $ (set $ r = 1 $ and $ m=t $ in Corollary \ref{cor_singleton}). Isolating $ d $, we obtain the relaxed Singleton bound
\begin{equation}
d \leq n - \left\lceil \frac{\kappa}{t} \right\rceil + 1.
\label{eq_Singleton_relaxed}
\end{equation}

\begin{definition} [\textbf{QMSRD codes} \cite{qmsrd}]
For an $ \F_q $-linear code $ \C \subseteq \F_q^{t \times n} $ as above, we say that it is quasi MSRD (or QMSRD) if it attains the relaxed Singleton bound (\ref{eq_Singleton_relaxed}). 
\end{definition}

For $ n_1 = n_2 = \ldots = n_\ell = 1 $ and arbitrary $ t,\ell \in \Z_+ $, QMSRD codes become quasi MDS (QMDS) codes \cite{qmds} (independently introduced as fractional MDS codes in \cite{ball2024griesmer} and additive MDS codes in \cite{yadav}). For $ \ell = 1 $ and all the other parameters being arbitrary, QMSRD codes become quasi MRD (QMRD) codes \cite{gorla}.

Notice that $ \C $ is MSRD if and only if it is QMSRD and $ t | \kappa $. In this case, the dual of an MSRD code is again MSRD (Theorem \ref{th_mosrd_duality} for $ r=1 $). However, the dual of a QMSRD code is not necessarily QMSRD if $ t \not| \kappa $ (see \cite{gorla, qmds, qmsrd, yadav}). This motivates the following.

\begin{definition} [\textbf{Dually QMSRD codes} \cite{qmsrd}]
For an $ \F_q $-linear code $ \C \subseteq \F_q^{t \times n} $ as above, we say that it is dually QMSRD if both $ \C $ and $ \C^\perp $ are QMSRD, where duals are taken with respect to the usual $ \F_q $-bilinear inner product in $ \F_q^{t \times n} $ (take $ m=1 $ and $ r=t $ in (\ref{eq_def_inner_product})).
\end{definition}

We will show in the following that MOSRD codes are dually QMSRD. However, we need to connect duality for $ \F_q $-linear codes and $ \F_{q^m} $-linear codes. The proof of the following lemma is straightforward.

\begin{lemma} \label{lemma_inner_products}
Let $ \aalpha =  (\alpha_1,\alpha_2, \ldots, \alpha_m) \in \F_{q^m}^m $ and $ \ggamma = (\gamma_1, \gamma_2, \ldots, \gamma_m) \in \mathbb{F}_{q^m}^m $ be ordered bases of $ \F_{q^m} $ over $ \F_q $ that are dual of each other (i.e., $ {\rm Tr}_q(\alpha_i \gamma_j) = \delta_{i,j} $, where $ {\rm Tr}_q $ denotes the trace of $ \F_{q^m} $ over $ \F_q $), see \cite[p. 54]{lidl}. Then for $ X = (X_1,X_2, \ldots, X_\ell),Y =(Y_1,Y_2, \ldots, Y_\ell) \in \mathbb{F}_{q^m}^{r \times n} $, we have
$$ {\rm Tr}_q(\langle X , Y \rangle) = \langle M_{\aalpha}(X) , M_{\ggamma}(Y) \rangle \in \F_q , $$
where the inner products are as in (\ref{eq_def_inner_product}) for the right parameters in each side of the equation (see Definition \ref{def_matrix_representation} for $ M_{\aalpha} $ and $ M_{\ggamma} $). In particular, for an $ \F_{q^m} $-linear code $ \C \subseteq \mathbb{F}_{q^m}^{r \times n} $, we have
$$ M_{\aalpha}(\mathcal{C})^\perp = M_{\ggamma}( \mathcal{C}^\perp). $$
\end{lemma}

In particular, since $ M_{\aalpha} : \F_{q^m}^{r \times n} \longrightarrow \F_q^{(mr) \times n} $ is an $ \F_q $-linear vector space isomorphism that preserves ranks, we may say that an $ \F_{q^m} $-linear code $ \C \subseteq \F_{q^m}^{r \times n} $ with $ k = \dim_{\F_{q^m}}(\C) $ is QMSRD if 
$$ \dd_{SR}(\C) = n - \left\lceil \frac{k}{r} \right\rceil + 1 . $$
Similarly, we may say that $ \C $ is dually QMSRD if both $ \C $ and $ \C^\perp $ (with respect to the $ \F_{q^m} $-bilinear inner product in $ \F_{q^m}^{r \times n} $) are QMSRD.

In order to connect MOSRD and dually QMSRD codes, we will make use of the following inequalities between the ordered sum-rank metric and the classical sum-rank metric.

\begin{lemma} \label{lemma_wtSR_wtOSR}
For $ X = (X_1,X_2, \ldots, X_\ell) \in \F_{q^m}^{r \times n} $, we have that
$$ \wt_{SR}(X) \leq \wt_{SR}^+(X) \leq r \cdot \wt_{SR}(X), $$
where $ \wt_{SR}(X) = \sum_{i=1}^\ell \wt_R(X_i) $ and $ \wt_R(X_i) $ is as in (\ref{eq_def_wtR}), for $ i \in [\ell] $. Similarly for $ \wt_{SR}^-(X) $.
\end{lemma}
\begin{proof}
By Definition \ref{def_OSR}, since $ X_i = X_i^{[r]} \in \F_{q^m}^{r \times n_i} $, for $ i \in [\ell] $, we have
$$ \sum_{i=1}^\ell \wt_R(X_i) \leq \sum_{i=1}^\ell \sum_{j=1}^r \wt_R(X_i^{[j]}) \leq r \sum_{i=1}^\ell \wt_R(X_i). $$
\end{proof}

Thus we can finally obtain the desired connection.

\begin{theorem} [\textbf{MOSRD and QMSRD}]
If an $ \F_{q^m} $-linear code $ \C \subseteq \F_{q^m}^{r \times n} $ is MOSRD (for any orientation), then it is dually QMSRD. In particular, if $ k = \dim_{\F_{q^m}}(\C) $ and $ r | k $, then $ \C $ is MSRD (and therefore MDS for the Hamming metric with alphabet $ \F_{q^m}^r $ and length $ n $).
\end{theorem}
\begin{proof}
Let $ k = \dim_{\F_{q^m}}(\C) $, $ d^+ = \dd_{SR}^+(\C) $ and $ d = \dd_{SR}(\C) $. Thus we have $ d^+ = rn - k + 1 $. By Lemma \ref{lemma_wtSR_wtOSR}, 
$$ d \geq \left\lceil \frac{d^+}{r} \right\rceil = \left\lceil \frac{rn - k + 1}{r} \right\rceil = n - \left\lceil \frac{k}{r} \right\rceil + 1. $$
Hence $ \C $ is QMSRD. Now by Theorem \ref{th_mosrd_duality}, we have that $ \C^\perp $ is negatively MOSRD, and we deduce similarly that $ \C^\perp $ is QMSRD, i.e., $ \C $ is dually QMSRD.
\end{proof}

In particular, we obtain the following.

\begin{corollary} \label{cor_LMC_QMSRD}
With notation and assumptions as in Definition \ref{def_lin_mult_codes}, for $ k \in [rn] $, the linearized multiplicity code $ \M_k(\aaa,\bbeta) \subseteq \F_{q^m}^{r \times n} $ is dually QMSRD of dimension $ k $ (over $ \F_{q^m} $). In particular, it is MSRD if $ r | k $ (and therefore MDS for the Hamming metric with alphabet $ \F_{q^m}^r $ and length $ n $).
\end{corollary}

The previous corollary recovers (by setting $ m=n_1=n_2= \ldots =n_\ell = 1 $) the fact that classical multiplicity codes are dually QMDS for the folded Hamming metric \cite[Cor. 31]{qmds} and they are MDS if $ r | k $ (which was implicitly shown earlier in \cite{multiplicity,nrt}). For $ \ell = 1 $ but the remaining parameters being arbitrary, we obtain that linearized multiplicity codes for the ordered rank metric (Remarks \ref{remark_lin_mult_gabidulin} and \ref{remark_ordered_rank}) are dually QMRD \cite{gorla}.

\section{The weight distribution of MOSRD codes} \label{sec_weight}

In this section, we compute the (ordered sum-rank) support and weight distributions of MOSRD codes. Notice that MacWilliams identities for duality do not hold already for the classical sum-rank metric \cite[Ex. V.2]{fundamental} (i.e., the case $ r=1 $). We will only treat the positively ordered sum-rank metric. The negative one is analogous.

\begin{definition} [\textbf{Distributions}] \label{def_distributions}
For a (linear or nonlinear) code $ \C \subseteq \F_{q^m}^{r \times n} $, we define its (positively) ordered sum-rank support distribution as the tuple $ S(\C) = (S_\LL(\C))_{\LL \in \PP^+_{q,\nn,r}} $, where
$$ S_\LL(\C) = |\{ X \in \C : \supp^+(X) = \LL \}| , $$
for $ \LL \in \PP^+_{q,\nn,r} $. We also define the (positively) ordered sum-rank weight distribution of $ \C $ as the tuple $ A(\C) = (A_w(\C))_{w=0}^{rn} $, where, for $ w \in [0,rn] $,
$$ A_w(\C) = |\{ X \in \C : \wt_{SR}^+(X) = w \}| . $$
\end{definition}

If $ r = 1 $ and the other parameters are arbitrary, support and weight distributions for the ordered sum-rank metric recover those for the classical sum-rank metric \cite[Def. II.8]{fundamental}. If $ m = n_1= n_2 = \ldots = n_\ell = 1 $ and $ r \in \Z_+ $ is arbitrary, then we obtain support and weight distributions for the classical NRT metric, which were considered in \cite{barg_park, martin, skriganov}.

When $ r = m = n_1 = n_2 = \ldots = n_\ell = 1 $, we recover the support and weight distributions of codes in the classical Hamming metric.

We start by computing the sizes of shortened codes. If $ \C $ is linear, take $ X_0 = 0 $ in the following lemma.

\begin{lemma} [\textbf{Shortened codes}] \label{lemma_shortened_MOSRD}
Let $ \C \subseteq \F_{q^m}^{r \times n} $ be a (linear or nonlinear) MOSRD code with $ d = \dd^+_{SR}(\C) $. Fix $ X_0 \in \C $. For $ \LL \in \PP^+_{q,\nn,r} $ with $ \rk(\LL) = w $, we have
$$ |(\C - X_0) \cap \V_\LL^+| = \left\lbrace \begin{array}{ll}
1  & \textrm{ if } w < d, \\
q^{m(w-d+1)} & \textrm{ if } w \geq d.
\end{array} \right. $$
\end{lemma}
\begin{proof}
Since the sizes and minimum distances of $ \C $ and $ \C-X_0 $ are equal, we may assume that $ 0 \in \C $ by replacing $ \C $ by $ \C - X_0 $.

From the definitions, $ \C \cap \V_\LL^+ = \{ 0 \} $ if $ w < d $. Assume now that $ w \geq d $. Similarly to the proof of Theorem \ref{th_mosrd_duality}, it is easy to see that we can choose $ \LL_0 \in \PP^+_{q,\nn,r} $ such that $ \LL_0 \subseteq \LL $ and $ \rk(\LL_0) = d-1 $. Consider the natural projection onto the quotient
$$ \pi : \C \longrightarrow \F_{q^m}^{r \times n} / \V_{\LL_0}^+. $$ 
This map is injective by Proposition \ref{prop_supports_basic_props}, since $ \rk(\LL_0) = d-1 < d = \dd_{SR}^+(\C) $ (linearity of $ \C $ is not needed, only that the metric is invariant by translations). Since the domain and codomain of $ \pi $ have the same (finite) size, then $ \pi $ is a bijection. In particular, the restriction
$$ \pi : (\C \cap \V_{\LL}^+) \longrightarrow \V_{\LL}^+ / \V_{\LL_0}^+ $$
is also a bijection. Taking sizes, the result follows.
\end{proof}

We also need to consider the M{\"o}bius function on the poset $ \PP^+_{q,\nn,r} $ (with componentwise set inclusions). We use \cite[Sec. 3.7, Eq. (3.15)]{stanley} as the definition.

\begin{definition} [\textbf{M{\"o}bius function}] \label{def_mobius}
We define the m{\"o}bius function $ \mu : (\PP^+_{q,\nn,r})^2 \longrightarrow \Z $ by $ \mu(\K,\LL) = 0 $ if $ \K \not\subseteq \LL $, $ \mu(\K,\LL) = 1 $ if $ \K = \LL $, and if $ \K \subseteq \LL $,
$$ \mu(\K,\LL) = - \sum_{\K \subseteq \M \subsetneq \LL} \mu(\K, \M). $$
\end{definition}

The M{\"o}bius function is well defined by strong induction on the (finite) size of the interval $ [\K, \LL] = \{ \M \in \PP^+_{q,\nn,r} : \K \subseteq \M \subseteq \LL \} $. We next compute this function on the poset $ \PP^+_{q,\nn,r} $.

\begin{lemma} \label{lemma_mobius_explicit}
Let $ \LL = (\LL_{i,j})_{i=1,j=1}^{\ell,r} , \K = (\K_{i,j})_{i=1,j=1}^{\ell,r} \in \PP^+_{q,\nn,r} $ be such that $ \K \subseteq \LL $. Then $ \mu(\K,\LL) = 0 $ unless $ \LL_{i,j-1} \subseteq \K_{i,j} $, for all $ i \in [\ell] $ and $ j \in [r] $ (defining $ \LL_{i,0} = 0 $), in which case,
$$ \mu(\K,\LL) = \prod_{i=1}^\ell \prod_{j=1}^r (-1)^{c_{i,j}} q^{\binom{c_{i,j}}{2}}, $$
where $ c_{i,j} = \dim(\LL_{i,j}) - \dim(\K_{i,j}) $, for $ i \in [\ell] $ and $ j \in [r] $.
\end{lemma}
\begin{proof}
Define $ \nu : (\PP^+_{q,\nn,r})^2 \longrightarrow \Z $ by $ \nu(\K,\LL) = 0 $ unless $ \K \subseteq \LL $ and $ \LL_{i,j-1} \subseteq \K_{i,j} $, for all $ i \in [\ell] $ and $ j \in [r] $, in which case, define 
$$ \nu(\K,\LL) = \prod_{i=1}^\ell \prod_{j=1}^r (-1)^{c_{i,j}} q^{\binom{c_{i,j}}{2}}, $$
for $ \K,\LL \in \PP^+_{q,\nn,r} $ and $ c_{i,j} = \dim(\LL_{i,j}) - \dim(\K_{i,j}) $, for $ i \in [\ell] $ and $ j \in [r] $. To show that $ \nu = \mu $, it is enough to prove that
\begin{equation}
\sum_{\K \subseteq \M \subseteq \LL} \nu(\K,\M) = \delta_{\K, \LL},
\label{eq_nu_identity}
\end{equation}
for $ \K, \LL \in \PP^+_{q,\nn,r} $ with $ \K \subseteq \LL $. Take one such pair $ \K,\LL $. Let $ \M \in \PP^+_{q,\nn,r} $ with $ \K \subseteq \M \subseteq \LL $. By definition, $ \nu(\K,\M) = 0 $ unless 
\begin{equation}
\K_{i,j} \subseteq \M_{i,j} \subseteq \K_{i,j+1} \cap \LL_{i,j},
\label{eq_M_inclusions}
\end{equation}
for all $ i \in [\ell] $ and all $ j \in [r] $ (defining $ \K_{i,r+1} = \F_q^{n_i} $). Notice also that, if $ \M_{i,j} \subseteq \F_q^{n_i} $ are subspaces satisfying (\ref{eq_M_inclusions}), for $ i \in [\ell] $ and $ j \in [r] $, then automatically $ \M = (\M_{i,j})_{i=1,j=1}^{\ell,r} \in \PP^+_{q,\nn,r} $.

Writing $ \mathcal{H}_{i,j} = \K_{i,j+1} \cap \LL_{i,j} $ and $ t_{i,j} = \dim(\mathcal{H}_{i,j}) - \dim(\K_{i,j}) $, for $ i \in [\ell] $ and $ j \in [r] $, we deduce that
$$ \sum_{\K \subseteq \M \subseteq \LL}
\nu(\K,\M) =
\prod_{i=1}^\ell\prod_{j=1}^r
\left( \sum_{\K_{i,j} \subseteq \mathcal{N} \subseteq \mathcal{H}_{i,j}}
(-1)^{\dim(\mathcal{N}) - \dim(\K_{i,j})}
q^{\binom{\dim(\mathcal{N})-\dim(\K_{i,j})}{2}}
\right) =
\prod_{i=1}^{\ell}\prod_{j=1}^{r}S_{t_{i,j}}, $$
where we define, for $ t \in \N $,
$$ S_t = \sum_{c=0}^{t} {t\brack c}_q (-1)^c q^{\binom c2}, \quad \textrm{where, if } 1 \leq c \leq t, \quad {t\brack c}_q = \prod_{i=0}^{c-1} \frac{q^t-q^i}{q^c-q^i} $$
is the $ q $-ary Gaussian coefficient, which measures the number of $ c $-dimensional $ \F_q $-linear subspaces of $ \F_q^t $ (see \cite[Ch. 24]{van_course}). We define $ {t\brack c}_q = \binom{t}{c} = 0 $ if $ t < c $ and $ {t\brack 0}_q = 1 $. Thus clearly $ S_0 = 1 $. We next show that $ S_t = 0 $ if $ t > 0 $. Define the polynomial
$$ F_t(z) = \sum_{c=0}^{t} {t\brack c}_q q^{\binom c2}z^c \in \Z[z] . $$
Using the well-known identity (see \cite[Eq. (24.1)]{van_course})
$$ {t\brack c}_q = {t-1\brack c}_q + q^{t-c}{t-1\brack c-1}_q, $$
one can easily deduce that $ F_t(z) = (1+q^{t-1}z)F_{t-1}(z) $, for $ t \in \Z_+ $. Since $F_0(z)=1$, then by induction we can see that, for $ t > 0 $,
\begin{equation}
F_t(z) = \prod_{h=0}^{t-1}(1+q^hz).
\label{eq_F_t}
\end{equation}
Therefore $ S_t = F_t(-1) = \prod_{h=0}^{t-1}(1-q^h) = 0 $ if $ t > 0 $. Hence, in order to conclude Equation (\ref{eq_nu_identity}), we just need to show that $ t_{i,j} = 0 $, for all $ i \in [\ell] $ and all $ j \in [r] $, if and only if $ \K = \LL $. The only nontrivial implication is the direct one. Assume that $ t_{i,j} = 0 $, that is, $ \K_{i,j} = \K_{i,j+1} \cap \LL_{i,j} $, for all $ i \in [\ell] $ and all $ j \in [r] $. Fix $ i \in [\ell] $. For $ j=r $, we have $ \K_{i,r} = \LL_{i,r} $ (since $ \K_{i,r+1} = \F_q^{n_i} $). Now assume that $ j \leq r-1 $ and $ \K_{i,j+1} = \LL_{i,j+1} $. Then $ K_{i,j} = \K_{i,j+1} \cap \LL_{i,j} = \LL_{i,j+1} \cap \LL_{i,j} = \LL_{i,j} $. By induction, we conclude that $ \K = \LL $, and we are done.
\end{proof}

We will also need the following M{\"o}bius generating polynomial.

\begin{lemma} \label{lemma_P_z}
Fix $ \LL \in \PP^+_{q,\nn,r} $ and let $ P_\LL(z) = \sum_{\K \in \PP^+_{q,\nn,r}, \K \subseteq \LL} \mu(\K,\LL) z^{\rk(\LL) - \rk(\K)} \in \Z[z] $. Then
$$ P_\LL(z) = \prod_{i=1}^\ell \prod_{j=1}^r \prod_{h=0}^{a_{i,j}-1} (1- q^h z), $$
where $ a_{i,j} = \dim(\LL_{i,j}) - \dim(\LL_{i,j-1}) $, for $ i \in [\ell] $ and $ j \in [r] $ (defining $ \LL_{i,0} = 0 $).
\end{lemma}
\begin{proof}
By Lemma \ref{lemma_mobius_explicit}, if $ \K \in \PP^+_{q,\nn,r} $ and $ \mu(\K,\LL) \neq 0 $, then $ \LL_{i,j-1} \subseteq \K_{i,j} \subseteq \LL_{i,j} $, for $ i \in [\ell] $ and $ j \in [r] $. Conversely, if $ \K_{i,j} \subseteq \F_q^{n_i} $ are subspaces such that $ \LL_{i,j-1} \subseteq \K_{i,j} \subseteq \LL_{i,j} $, for $ i \in [\ell] $ and $ j \in [r] $, then $ \K = (\K_{i,j})_{i=1,j=1}^{\ell,r} \in \PP^+_{q,\nn,r} $. Hence, we may identify the tuples $ \K = (\K_{i,j})_{i=1,j=1}^{\ell,r} \in \PP^+_{q,\nn,r} $ such that $ \mu(\K,\LL) \neq 0 $ with tuples of arbitrary subspaces of the quotient $ \LL_{i,j} / \LL_{i,j-1} $, for $ i \in [\ell] $ and $ j \in [r] $. Since such a quotient has dimension $ a_{i,j} $, the number of such subspaces $ \K_{i,j} \subseteq \F_q^{n_i} $ with $ c = \dim(\LL_{i,j}/\K_{i,j}) $ is $ {a_{i,j}\brack a_{i,j} - c}_q = {a_{i,j}\brack c}_q $. Thus by Lemma \ref{lemma_mobius_explicit},
\begin{equation}
\begin{split}
P_{\mathcal L}(z) & = \prod_{i=1}^\ell \prod_{j=1}^r \left( \sum_{c=0}^{a_{i,j}} {a_{i,j}\brack c}_q (-1)^c q^{\binom c2} z^c \right) \\
& = \prod_{i=1}^\ell \prod_{j=1}^r \prod_{h=0}^{a_{i,j}-1}(1-q^hz),
\end{split}
\label{eq_P_L}
\end{equation}
where the last equality follows from Equation (\ref{eq_F_t}).
\end{proof}

We may now compute the support distribution of MOSRD codes. If $ \C $ is linear, take $ X_0 = 0 $ in the following lemma.

\begin{proposition} [\textbf{MOSRD support distribution}] \label{prop_MOSRD_support_dist}
Let $ \C \subseteq \F_{q^m}^{r \times n} $ be a (linear or nonlinear) MOSRD code with $ d = \dd_{SR}^+(\C) $. Fix $ X_0 \in \C $. For $ \LL \in \PP^+_{q,\nn,r} $ with $ w = \rk(\LL) $, let $ P_\LL(z) = \sum_{s \in \N} p_s(\LL) z^s \in \Z[z] $ be the polynomial from Lemma \ref{lemma_P_z}, where $ p_s(\LL) \in \Z $ for $ s \in \N $. Then $ S_\LL(\C - X_0) = 1 $ if $ w = 0 $, $ S_\LL(\C - X_0) = 0 $ if $ w \in [d-1] $, and if $ w \in [d,rn] $, then
$$ S_\LL(\C - X_0) = \sum_{s=0}^{w-d} \left( q^{m(w-d+1-s)} - 1 \right) p_s(\LL). $$
\end{proposition}
\begin{proof}
As in the proof of Lemma \ref{lemma_shortened_MOSRD}, we may assume that $ 0 \in \C $ by replacing $ \C $ by $ \C-X_0 $. Let $ \LL \in \PP^+_{q,\nn,r} $ with $ w = \rk(\LL) $. The cases $ w \in [0,d-1] $ are trivial, thus we assume $ w \in [d,rn] $. We clearly have that
$$ |\C \cap \V_\LL^+| = \sum_{\K \in \PP^+_{q,\nn,r}, \K \subseteq \LL} S_\K(\C). $$
By the M{\"o}bius inversion formula \cite[Prop. 3.7.1]{stanley}, we have that
$$ S_\LL(\C) = \sum_{\K \in \PP^+_{q,\nn,r}, \K \subseteq \LL} \mu(\K,\LL) |\C \cap \V_\K^+ |. $$
Now, if $ 0 $ denotes the tuple of zero subspaces in $ \PP^+_{q,\nn,r} $, then by Definition \ref{def_mobius}, $ \sum_{\K \in \PP^+_{q,\nn,r}, \K \subseteq \LL} \mu(\K,\LL) $ $ = 0 $. Together with Lemma \ref{lemma_shortened_MOSRD}, we deduce that
\begin{equation*}
\begin{split}
S_\LL(\C) & = \sum_{\K \in \PP^+_{q,\nn,r}, \K \subseteq \LL} \mu(\K,\LL) |\C \cap \V_\K^+ | \\
 & = \sum_{\substack{\K \in \PP^+_{q,\nn,r}, \\ \K \subseteq \LL, \rk(\K) \geq d }} \mu(\K,\LL) \left( q^{m(\rk(\K) - d + 1)} - 1 \right) \\
 & = \sum_{s=0}^{w-d} \left( \left( q^{m(w-d+1-s)} - 1 \right) \sum_{\substack{\K \in \PP^+_{q,\nn,r}, \\ \K \subseteq \LL, \rk(\K) = w - s }} \mu(\K,\LL) \right).
\end{split}
\end{equation*}
Thus the result follows from the definition of $ P_\LL(z) $ from Lemma \ref{lemma_P_z}.
\end{proof}

If $ r = 1 $ and the other parameters are arbitrary, Proposition \ref{prop_MOSRD_support_dist} becomes the known sum-rank support distribution of MSRD codes \cite[Th. VI.5]{fundamental}. If $ m=n_1=n_2= \ldots = n_\ell = 1 $, the support distribution of linear ordered MDS codes can be obtained from \cite[Prop. 2.10]{barg_park}.

Finally, we may compute the weight distribution of MOSRD codes. Define
$$ \A_{\nn,r} = \left\lbrace (a_{i,j})_{i=1,j=1}^{\ell,r+1} \in \N^{\ell (r+1)} : \sum_{j=1}^{r+1} a_{i,j} = n_i, \forall i \in [\ell] \right\rbrace, $$
and for $ \aaa \in \A_{\nn,r} $, define the polynomial
$$ P_\aaa (z) = \prod_{i=1}^\ell \prod_{j=1}^r \prod_{h=0}^{a_{i,j}-1} (1- q^h z) = \sum_{s \in \N} p_s(\aaa) z^s \in \Z[z], $$
where $ p_s(\aaa) \in \Z $, for all $ s \in \N $. For $ a_1,a_2, \ldots, a_{r+1} \in \N $ with $ \sum_{j=1}^{r+1} a_j = n $, we also define the $ q $-ary generalized Gaussian coefficient
$$ \qmultinom{n}{a_1,a_2,\ldots,a_{r+1}} = \frac{[n]_q !}{ [a_1]_q ! \cdot [a_2]_q ! \cdots [a_{r+1}]_q ! } \in \Z_+, \quad \textrm{ where } [t]_q ! = \prod_{h=1}^t \frac{q^h-1}{q-1}, $$
for $ t \in \Z_+ $, and $ [0]_q! = 1 $. The above $ q $-ary Gaussian coefficient counts the number of flags of subspaces
$$ \LL_1 \subseteq \LL_2 \subseteq \ldots \subseteq \LL_r \subseteq \F_q^n $$
such that $ a_j = \dim(\LL_j) - \dim(\LL_{j-1}) $, for $ j \in [r+1] $, where $ \LL_0 = 0 $ and $ \LL_{r+1} = \F_q^n $. See \cite[Sec. 1.7]{stanley}. If $ \C $ is linear, take $ X_0 = 0 $ in the following lemma.

\begin{theorem}[\textbf{MOSRD weight distribution}] \label{th_MOSRD_weight_dist}
Let $ \C \subseteq \F_{q^m}^{r \times n} $ be a (linear or nonlinear) MOSRD code with $ d = \dd_{SR}^+(\C) $. Fix $ X_0 \in \C $. Then $ A_0(\C - X_0) = 1 $, $ A_w(\C - X_0) = 0 $ for $ w \in [d-1] $, and 
$$ A_w(\C - X_0) = \sum_{\aaa \in \A_{\nn,r}, \omega(\aaa)=w} \left( \prod_{i=1}^\ell \qmultinom{n_i}{a_{i,1},\ldots,a_{i,r+1}} \right) \sum_{s=0}^{w-d} \left( q^{m(w-d+1-s)} - 1 \right)p_s(\aaa) $$
for $ w \in [d,rn] $, where $ \omega(\aaa) = \sum_{i=1}^\ell \sum_{j=1}^r a_{i,j}(r-j+1) $, for $ \aaa = (a_{i,j})_{i=1,j=1}^{\ell,r+1} \in \A_{\nn,r} $.
\end{theorem}
\begin{proof}
As in the proof of Lemma \ref{lemma_shortened_MOSRD}, we may assume that $ 0 \in \C $ by replacing $ \C $ by $ \C - X_0 $. Again, the cases $ w \in [0,d-1] $ are trivial, thus we assume that $ w \in [d,rn] $. Using Proposition \ref{prop_MOSRD_support_dist}, we deduce that
\begin{equation*}
\begin{split}
A_w (\C) & = \sum_{\LL \in \PP^+_{q,\nn,r}, \rk(\LL) = w} S_\LL(\C) \\
 & = \sum_{\LL \in \PP^+_{q,\nn,r}, \rk(\LL) = w} \sum_{s=0}^{w-d} \left( q^{m(w-d+1-s)} - 1 \right) p_s(\LL) \\
 & = \sum_{\aaa \in \A, \omega(\aaa)=w} \left( \prod_{i=1}^\ell \qmultinom{n_i}{a_{i,1},\ldots,a_{i,r+1}} \right) \sum_{s=0}^{w-d} \left( q^{m(w-d+1-s)} - 1 \right)p_s(\aaa) ,
\end{split}
\end{equation*}
where the last equality follows from the discussion on $ q $-ary generalized Gaussian coefficients above this theorem, and from the fact that $ p_s(\LL) = p_s(\aaa) $, where $ a_{i,j} = \dim(\LL_{i,j}) - \dim(\LL_{i,j-1}) $, for $ i \in [\ell] $ and $ j \in [r] $. 
\end{proof}

If $ r = 1 $ and the other parameters are arbitrary, then Theorem \ref{th_MOSRD_weight_dist} becomes the known sum-rank weight distribution of MSRD codes \cite[Remark VI.11]{fundamental}. If $ m = n_1= n_2 = \ldots = n_\ell = 1 $ and $ r \in \Z_+ $ is arbitrary, then Theorem \ref{th_MOSRD_weight_dist} becomes the known NRT weight distribution of ordered MDS codes \cite[Th. 3.1]{skriganov}. 

When $ r = m = n_1 = n_2 = \ldots = n_\ell = 1 $, Theorem \ref{th_MOSRD_weight_dist} recovers the Hamming weight distribution of classical MDS codes \cite[Th. 7.4.1]{pless}.

\section{The extension degree $ m $ vs the column numbers $ n_i $} \label{sec_upper_bounds_m}

In the classical sum-rank metric (when $ r = 1 $), one can obtain a second Singleton bound by transposing matrices in Corollary \ref{cor_singleton}, see \cite[Cor. 3]{universal-lrc} or \cite[Th. 1.5]{fnt}. From this bound one can deduce that, when $ \tilde{n} = n_1 = n_2 = \ldots = n_\ell $, the extension degree satisfies $ m \geq \tilde{n} $. 

Such a simple transposition technique does not seem to work for the ordered sum-rank metric for $ r \geq 2 $. However, in this section, we still prove $ m \geq \tilde{n} $ for almost all cases, which is attained by linearized multiplicity codes. That is, they are extremal MOSRD codes for $ m $ and $ \tilde{n} $.

We first prove the following lemma on minimal codewords of MOSRD codes.

\begin{lemma} \label{lemma_support_for_m_bound}
Let $ \C \subseteq \F_{q^m}^{r \times n} $ be an $ \F_{q^m} $-linear MOSRD code and let $ \LL \in \PP^+_{q,\nn,r} $ with $ \rk(\LL) = \dd_{SR}^+(\C) $. Then there exists $ X \in \C $ such that $ \supp^+(X) = \LL $. In fact, such codewords, together with the zero codeword, form a one-dimensional $ \F_{q^m} $-linear subcode of $ \C $.
\end{lemma}
\begin{proof}
By Lemma \ref{lemma_shortened_MOSRD}, $ \dim_{\F_{q^m}}(\C \cap \V_{\LL}^+) = \rk(\LL) - \dd_{SR}^+(\C) + 1 = 1 $. Now, for any nonzero $ X \in \C \cap \V_\LL^+ $, we have that
$$ \dd_{SR}^+(\C) \leq \wt_{SR}^+(X) = \rk(\supp^+(X)) \leq \rk(\LL) = \dd_{SR}^+(\C), $$
by Proposition \ref{prop_supports_basic_props}, and we are done.
\end{proof}

\begin{theorem}[\textbf{Bounding $ m $ by $ \tilde{n} $}]
Assume $ n_1 = n_2 = \ldots = n_\ell $, and denote this number by $ \tilde{n} $. Assume there exists an $ \F_{q^m} $-linear MOSRD code $ \C \subseteq \F_{q^m}^{r \times n} $ (where $ n = \ell \tilde{n} $). Let $ d = \dd_{SR}^+(\C) $ and $ d^\perp = \dd_{SR}^-(\C^\perp) $, and assume that $ d \in [2,rn] $ (i.e., $ \C $ is not $ \{ 0 \} $ or $ \F_{q^m}^{r \times n} $, thus also $ d^\perp \in [2,rn] $). We have the following bounds:
\begin{alignat}{2}
m & \geq \tilde{n}
&\qquad\qquad&
\text{if } \ell \geq 2 \textrm{ or } r = 1 \textrm{ or } \tilde{n} | \max \{d,d^\perp\} ,
\label{eq_bound_m_ell_geq_2}\\
2m &\geq \tilde{n}
&&
\text{otherwise}.
\label{eq_bound_m_ell=1}
\end{alignat}
In particular, the bound (\ref{eq_bound_m_ell_geq_2}) is sharp since linearized multiplicity codes (Definition \ref{def_lin_mult_codes}) attain it in such a parameter regime (see their restrictions in Equation (\ref{eq_lin_mult_code_restrictions})).
\end{theorem}
\begin{proof}
First, the case $ r = 1 $ was proven in \cite[Cor. 3]{universal-lrc}. See also \cite[Prop. 1.8]{fnt}.

By Theorem \ref{th_mosrd_duality}, $ d + d^\perp = \ell r \tilde{n} + 2 $. Without loss of generality, we may assume that $ d \geq d^\perp $ and prove the result for the positively ordered sum-rank metric. 

Assume first $ \ell \geq 2 $. From $ d + d^\perp = \ell r \tilde{n} + 2 \geq 2 (r \tilde{n} + 1) $ and $ d \geq d^\perp $, we deduce that $ d \geq r \tilde{n} + 1 $. Write
$$ d = (r \tilde{n}) s + \delta, \quad \textrm{where} \quad s,\delta \in \N \textrm{ and } 0 \leq \delta < r \tilde{n}.  $$
Notice that $ 1 \leq s \leq \ell $ since $ r \tilde{n} + 1 \leq d \leq \ell r \tilde{n} $. We next construct $ \LL = (\LL_{i,j})_{i=1,j=1}^{\ell,r} \in \PP^+_{q,\nn,r} $ with $ \rk(\LL) = d $. For $ i \in [s] $, define
$$ \LL_{i,1} = \LL_{i,2} = \ldots = \LL_{i,r} = \F_q^{\tilde{n}}. $$ 
If $ \delta = 0 $, define $ \LL_{i,j} = 0 $ for all $ i \in [s+1,\ell] $ and $ j \in [r] $. If $ \delta > 0 $, write
$$ \delta = \tilde{n} t + \partial, \quad \textrm{where} \quad t,\partial \in \N \textrm{ and } 0 \leq \partial < \tilde{n}, $$
where $ 0 \leq t \leq r $ since $ 0 \leq \delta < r \tilde{n} $. Now choose an $ \F_q $-linear subspace $ \mathcal{U} \subseteq \F_q^{\tilde{n}} $ with $ \dim_{\F_q}(\mathcal{U}) = \partial $ and define
$$ \LL_{s+1,j} = \left\lbrace \begin{array}{ll}
0 & \textrm{ if } j < r - t, \\
\mathcal{U} & \textrm{ if } j = r-t, \\
\F_q^{\tilde{n}} & \textrm{ if } j > r - t,
\end{array} \right. $$
for $ j \in [r] $. Finally, define $ \LL_{i,j} = 0 $ for all $ i \in [s+2,\ell] $ and $ j \in [r] $. Thus in all cases ($ \delta = 0 $ or $ \delta > 0 $) we have
$$ \rk(\LL) = (r \tilde{n}) s + \delta = d. $$
By Lemma \ref{lemma_support_for_m_bound}, there exists a nonzero $ X = (X_1,X_2, \ldots, X_\ell) \in \C $ such that $ \supp^+(X) = \LL $. Now, fixing $ i \in [\ell] $, with the decomposition of $ X $ as in Definition \ref{def_OSR}, we have that
$$ \supp^+_{i,1}(X) = \row(M_{\aalpha}(X_i^{[1]})) = \row(M_{\aalpha}(\xx_{i,1})), $$
for some ordered basis $ \aalpha \in \F_{q^m}^m $ of $ \F_{q^m} $ over $ \F_q $. Since $ M_{\aalpha}(\xx_{i,1}) \in \F_q^{m \times \tilde{n}} $, we deduce that
$$ \dim_{\F_q}(\supp^+_{i,1}(X)) \leq m. $$ 
However, taking $ i = 1 \in [s] $ (recall that $ s \geq 1 $), then 
$$ \supp^+_{1,1}(X) = \LL_{1,1} = \F_q^{\tilde{n}}, $$
and we conclude that $ \tilde{n} \leq m $.

Assume now that $ \ell = 1 $ and $ r \geq 2 $. In this case, $ d + d^\perp = r \tilde{n} + 2 \geq 2 (\tilde{n} + 1) $ and $ d \geq d^\perp $, thus we deduce that $ d \geq \tilde{n} + 1 $. As above, write
$$ d = \tilde{n} s + \delta, \quad \textrm{where} \quad s,\delta \in \N \textrm{ and } 0 \leq \delta < \tilde{n}.  $$
Notice that $ 1 \leq s \leq r $ since $ \tilde{n} + 1 \leq d \leq r \tilde{n} $, and if $ s = r $, then necessarily $ \delta = 0 $. Let $ \mathcal{U} \subseteq \F_q^{\tilde{n}} $ be an $ \F_q $-linear subspace with $ \dim_{\F_q}(\mathcal{U}) = \delta $, and define
$$ \LL_{1,j} = \left\lbrace \begin{array}{ll}
0 & \textrm{ if } j < r - s, \\
\mathcal{U} & \textrm{ if } j = r-s, \\
\F_q^{\tilde{n}} & \textrm{ if } j > r - s,
\end{array} \right. $$
for $ j \in [r] $. If $ \LL = (\LL_{1,j})_{j=1}^r $ (recall that $ \ell = 1 $), then $ \rk(\LL) = \tilde{n} s + \delta = d $. As above, by Lemma \ref{lemma_support_for_m_bound}, there exists a nonzero $ X \in \C $ such that $ \supp^+(X) = \LL $. As above, we have that
$$ \dim_{\F_q}(\LL_{1,j}) - \dim_{\F_q}(\LL_{1,j-1}) \leq \rk(M_{\aalpha}(\xx_{1,j})) \leq m, $$
for $ j \in [r] $ (defining $ \LL_{1,0} = 0 $), since $ M_{\aalpha}(\xx_{1,j}) \in \F_q^{m \times \tilde{n}} $. Now if $ \delta = 0 $ (i.e., $ \tilde{n} | d $), then
$$ \dim_{\F_q}(\LL_{1,r-s+1}) - \dim_{\F_q}(\LL_{1,r-s}) = \tilde{n} \leq m, $$
and if $ \delta > 0 $, then
$$ \dim_{\F_q}(\LL_{1,r-s+1}) - \dim_{\F_q}(\LL_{1,r-s-1}) = \delta + (\tilde{n} - \delta) = \tilde{n} \leq 2m. $$
\end{proof}

For $ \ell = 1 $, we may find an example with $ \tilde{n} = 2m $, i.e., the bound $ \tilde{n} \leq m $ is not always true. 

\begin{example}
Let $ \ell = m = 1 $ and $ r = \tilde{n} = n_1 = 2 $, i.e., $ n = \ell \tilde{n} = 2 $. The ambient space is $ \F_q^{2 \times 2} $. Let $ \aalpha = (\alpha_1,\alpha_2) \in \F_{q^2}^2 $ be an ordered basis of $ \F_{q^2} $ over $ \F_q $ and define
$$ \C = \{ (M_{\aalpha}(a \alpha_1),M_{\aalpha}(a \alpha_2)) : a \in \F_{q^2} \} \subseteq \F_q^{2 \times 2}. $$
Since all nonzero matrices of $ \C $ are invertible, we deduce that $ d = d_{SR}^+(\C) = 3 $. Since $ k = \dim_{\F_q}(\C) = \dim_{\F_q}(\F_{q^2}) = 2 $, we conclude that $ \C $ is MOSRD, and $ \tilde{n} = 2 = 2 m $. 
\end{example}

\section{The length $ \ell $ vs the field size $ q $} \label{sec_upper_bounds_ell}

In this section, we provide upper bounds on the block length $ \ell $ by the field size $ q $ for $ \F_{q^m} $-linear MOSRD codes with square matrices ($ m = n_1 = n_2= \ldots = n_\ell $). In all cases when $ r \geq 2 $ and $ m \geq 3 $, we will show that linearized multiplicity codes attain exactly the maximum length $ \ell $ relative to $ q $. In the other cases, they attain the same growth order. That is, they are extremal MOSRD codes for $ \ell $ and $ q $.

In the classical Hamming metric, it is well-known that from the weight distribution of MDS codes \cite[Th. 7.4.1]{pless} one can obtain the upper bound
\begin{equation}
n \leq 2q-2
\label{eq_2q-2_MDS}
\end{equation}
for the length $ n $ of a linear MDS code of dimension $ k \in [2,n-2] $ over an alphabet of size $ q $, see \cite[Cor. 7.4.4]{pless}. MDS codes are MOSRD codes for $ r = m = n_1=n_2= \ldots = n_\ell=1 $, thus the block length is $ n = \ell $. 

In principle, $ n \leq 2q-2 $ looks much worse than the MDS conjecture $ n \leq q+1 $ (proven for prime $ q $ in \cite{ball}). However, it shows that MDS codes must satisfy $ n = \mathcal{O}(q) $, which is attained by Reed--Solomon codes \cite{reed-solomon}. Furthermore, for $ q $ even, the MDS conjecture on $ \log_2(n) $ only improves by one additive unit the bound (\ref{eq_2q-2_MDS}). Thus in practice (\ref{eq_2q-2_MDS}) is already an important bound.

As in Section \ref{sec_weight}, we only treat the positively ordered sum-rank metric. The negative one is analogous. We start with the following technical bound from support distributions. As usual, we define $ [a]_q = {a \brack 1}_q = \frac{q^a-1}{q-1} $ for $ a \in \Z_+ $ and $ [0]_q = 0 $.

\begin{lemma} \label{lemma_lowerbound_qm+1}
Let $ \C \subseteq \F_{q^m}^{r \times n} $ be an $ \F_{q^m} $-linear MOSRD code with $ d = \dd_{SR}^+(\C) $. Let $ \LL \in \PP^+_{q,\nn,r} $ such that $ \rk(\LL) = d+1 $ and $ a_{i,j} = \dim(\LL_{i,j}) - \dim(\LL_{i,j-1}) $, for $ i \in [\ell] $ and $ j \in [r] $ (defining $ \LL_{i,0} = 0 $). Then
$$ -p_1(\LL) = \sum_{i=1}^\ell \sum_{j=1}^r [a_{i,j}]_q \leq q^m + 1, $$
where $ P_\LL(z) = \sum_{s \in \N} p_s(\LL) z^s \in \Z[z] $ is the polynomial from Lemma \ref{lemma_P_z}, where $ p_s(\LL) \in \Z $ for $ s \in \N $.
\end{lemma}
\begin{proof}
The only nonzero terms of $ S_\LL(\C) $ in Proposition \ref{prop_MOSRD_support_dist} correspond to $ p_0(\LL) $ and $ p_1(\LL) $ since $ \rk(\LL) = d+1 $. By Equation (\ref{eq_P_L}), we have that $ p_0(\LL) = 1 $ and
$$ p_1(\LL) = - \sum_{i=1}^\ell \sum_{j=1}^r [a_{i,j}]_q . $$
Therefore, by Proposition \ref{prop_MOSRD_support_dist}, we have
$$ S_\LL(\C) = (q^{2m}-1)p_0(\LL) + (q^m-1)p_1(\LL) = (q^m-1) \left( q^m + 1 - \sum_{i=1}^\ell \sum_{j=1}^r [a_{i,j}]_q \right). $$
The bound now follows from the fact that $ S_\LL(\C) \geq 0 $.
\end{proof}

We may deduce the following bounds on $ \ell $ by $ q $ for square matrices. In contrast with the classical Hamming metric ($ r=m=n_1=n_2= \ldots = n_\ell=1 $), we will need the finer support distribution, instead of the weight distribution, of MOSRD codes.

\begin{theorem} [\textbf{Bounding $ \ell $ by $ q $}] \label{th_bounds_ell_q}
Assume $ m = n_1 = n_2 = \ldots = n_\ell $. Assume there exists an $ \F_{q^m} $-linear MOSRD code $ \C \subseteq \F_{q^m}^{r \times n} $ with $ k = \dim_{\F_{q^m}}(\C) $ and
\begin{equation}
2 \leq k \leq r \ell m - 2.
\label{eq_assumption_on_k}
\end{equation}
We have the following bounds:
\begin{alignat}{2}
\ell &\leq q+1
&\qquad\qquad&
\text{if } r \geq 2 \text{ and } m=1,
\label{eq_bound_m=1}\\
\ell &\leq q
&&
\text{if } r \geq 2, m=2 \textrm{ and } k \textrm{ is odd},
\label{eq_bound_m=2}\\
\ell &\leq q-1
&&
\text{if } r \geq 2, \text{ and } m\geq 3 \textrm{ or (} m=2 \textrm{ and } k \textrm{ is even)},
\label{eq_bound_mgeq3}\\
\ell &\leq 2q-3
&&
\text{if } r=1 \text{ and } m\geq 2,
\label{eq_bound_r=1}\\
\ell &\leq 2q-2
&&
\text{if } r=m=1.
\label{eq_bound_mds}
\end{alignat}
In particular, the bound (\ref{eq_bound_mgeq3}) is sharp since linearized multiplicity codes (Definition \ref{def_lin_mult_codes}) attain it in such a parameter regime (see their restrictions in Equation (\ref{eq_lin_mult_code_restrictions})). Asymptotically, all of the bounds above read $ \ell = \mathcal{O}(q) $, an order of growth attained by linearized multiplicity codes in all cases.
\end{theorem}
\begin{proof}
First, (\ref{eq_bound_mds}) is the well-known bound (\ref{eq_2q-2_MDS}) for classical MDS codes. Now we prove the other ones. Let $ N = r \ell m $, $ d = N - k + 1 $ and $ d^\perp = k + 1 $ ($ d^\perp $ is the minimum distance of $ \C^\perp $, for the negative order, by Theorem \ref{th_mosrd_duality}).

Assume first $ r \geq 2 $. We have that
$$ (d+1) + (d^\perp + 1) = N+4 = r \ell m + 4 . $$
Since $ r \geq 2 $, then $ d+1 \geq \ell m +2 $ or $ d^\perp+1 \geq \ell m +2 $. We may assume the first of the two without loss of generality. By the assumptions (\ref{eq_assumption_on_k}), we have $ d+1 \leq N $. Write 
$$ d+1 = m s + \delta , \quad \textrm{ where } \quad s, \delta \in \N \textrm{ and } 0 \leq \delta < m . $$
Thus $ \ell \leq s \leq r \ell $. Choose $ s_1,s_2, \ldots, s_\ell \in [r] $ such that $ s = s_1 + s_2 + \cdots + s_\ell $. 

If $ \delta = 0 $, let $ \LL = (\LL_{i,j})_{i=1,j=1}^{\ell,r} \in \PP^+_{q,\nn,r} $ be given as follows. Define $ j_i = r - s_i + 1 \in [r] $ and 
\begin{equation}
\LL_{i,j} = \left\lbrace \begin{array}{ll}
0 & \textrm{ if } j < j_i, \\
\F_q^m & \textrm{ if } j \geq j_i,
\end{array} \right.
\label{eq_def_L_ij_bound_ell}
\end{equation}
for $ i \in [\ell] $. If $ a_{i,j} = \dim(\LL_{i,j}) - \dim(\LL_{i,j-1}) $ for $ i \in [\ell] $ and $ j \in [r] $ (defining $ \LL_{i,0} = 0 $), then $ a_{i,j_i} = m $ and $ a_{i,j} = 0 $ if $ j \neq j_i $, for $ i \in [\ell] $. In particular,
$$ -p_1(\LL) = \sum_{i=1}^\ell \sum_{j=1}^r [a_{i,j}]_q = \ell [m]_q , \quad \textrm{ and} $$
$$ \rk(\LL) = \sum_{i=1}^\ell \sum_{j=1}^r \dim(\LL_{i,j}) = \sum_{i=1}^\ell (r-j_i+1)m = \sum_{i=1}^\ell s_im = sm = d+1 . $$
If $ \delta > 0 $, then $ s < r \ell $ and we can choose $ s_1 \leq r-1 $. Thus if we define $ j_i = r - s_i + 1 \in [r] $ as above, for $ i \in [\ell] $, then $ j_1 \geq 2 $. Now let $ \LL = (\LL_{i,j})_{i=1,j=1}^{\ell,r} \in \PP^+_{q,\nn,r} $ be given as in Equation (\ref{eq_def_L_ij_bound_ell}) for $ i \in [2,\ell] $, and define
$$ \LL_{1,j} = \left\lbrace \begin{array}{ll}
0 & \textrm{ if } j < j_1 - 1, \\
\mathcal{U} & \textrm{ if } j = j_1 - 1, \\
\F_q^m & \textrm{ if } j \geq j_1,
\end{array} \right. $$
where $ \mathcal{U} \subseteq \F_q^m $ is an $ \F_q $-linear subspace of dimension $ \delta $. Now $ a_{i,j} = \dim(\LL_{i,j}) - \dim(\LL_{i,j-1}) $ for $ i \in [\ell] $ and $ j \in [r] $, are as above if $ i \geq 2 $, but 
$$ a_{1, j_1-1} = \delta \quad \textrm{and} \quad a_{1,j_1} = m - \delta . $$
Therefore, we now have that
$$ -p_1(\LL) = \sum_{i=1}^\ell \sum_{j=1}^r [a_{i,j}]_q = (\ell - 1) [m]_q + [m-\delta]_q + [\delta]_q , \quad \textrm{ and} $$
$$ \rk(\LL) = \sum_{i=1}^\ell \sum_{j=1}^r \dim(\LL_{i,j}) = \sum_{i=1}^\ell (r-j_i+1)m + \delta = \sum_{i=1}^\ell s_im + \delta = sm + \delta = d+1 . $$
By Lemma \ref{lemma_lowerbound_qm+1}, we have $ -p_1(\LL) \leq q^m+1 $.

If $ m = 1 $, then $ \delta = 0 $ and we obtain $ - p_1(\LL) = \ell \leq q+1 $ since $ [1]_q = 1 $ in this case.

Assume next that $ m \geq 2 $. If $ \ell \geq q $ and $ \delta = 0 $, then
$$ -p_1(\LL) = \ell [m]_q \geq q [m]_q = q \frac{q^m-1}{q-1} > q^m + 1, $$
which contradicts $ -p_1(\LL) \leq q^m+1 $, thus $ \ell \leq q-1 $ in this case. Assume now that $ \ell \geq q $ and $ 0 < \delta < m $. If $ m \geq 3 $, then $ \delta \geq 2 $ or $ m-\delta \geq 2 $, and thus
$$ - p_1(\LL) = (\ell - 1 )[m]_q + [\delta]_q + [m-\delta]_q \geq (q^m-1) + (q+1) + 1 \geq q^m + 2 , $$
which again contradicts $ -p_1(\LL) \leq q^m + 1 $, implying $ \ell \leq q-1 $ in this case. If $ m = 2 $, then $ 0 < \delta < m $ implies $ \delta = 1 $. Assume that $ \ell \geq q+1 $. Then
$$ -p_1(\LL) = (\ell - 1) [2]_q + 2 [1]_q > (q^2 -1) + 2 = q^2 + 1 , $$
contradicting $ -p_1(\LL) \leq q^2+1 $, hence $ \ell \leq q $ in this case. Assume next that $ k $ is even (and $ m = 2 $). Notice that $ N = 2 r \ell $ is even, thus so are $ d+1 $ and $ d^\perp + 1 $. Thus in this case it must be $ \delta = 0 $, in which case we have already proven that $ \ell \leq q-1 $.

Finally, we treat the case $ r=1 $ and $ m \geq 2 $. Again, we write
$$ d+1 = m s + \delta , \quad \textrm{ where } \quad s, \delta \in \N \textrm{ and } 0 \leq \delta < m . $$
In this case, we may choose $ \LL = (\LL_{i,1})_{i=1}^\ell \in \PP^+_{q,\nn,1} $ given by 
$$ \LL_{i,1} = \left\lbrace \begin{array}{ll}
\F_q^m & \textrm{ if } 1 \leq i \leq s, \\
\mathcal{U} & \textrm{ if } i = s+1 \textrm{ and } \delta > 0, \\
0 & \textrm{ otherwise} ,
\end{array} \right. $$
where $ \mathcal{U} \subseteq \F_q^m $ is an $ \F_q $-linear subspace of dimension $ \delta $. Thus in this case, by Lemma \ref{lemma_lowerbound_qm+1},
$$ -p_1(\LL) = s [m]_q + [\delta]_q \leq q^m + 1, $$
where $ [0]_q = 0 $. This inequality implies that $ s \leq q-1 $, and if $ s=q-1 $ then $ \delta \leq 1 $. Using the same argument to the dual distance $ d^\perp $, we obtain
$$ d \leq m(q-1) \quad \textrm{and} \quad d^\perp \leq m(q-1). $$
Using that $ d+d^\perp = \ell m + 2 $, we obtain $ \ell m + 2 \leq 2 m (q-1) $. Using that $ \ell $ is an integer and $ m \geq 2 $, we conclude that $ \ell \leq 2q - 3 $.
\end{proof}

The bound (\ref{eq_bound_r=1}) for MSRD codes ($ r = 1 $) already follows from the support distributions in \cite{fundamental}. For $ m = 1 $ and arbitrary $ r \geq 2 $ (but restricted $ k $), the bound $ \ell \leq q+1 $ appeared in \cite[Cor. 4.21]{Niederreiter1992} (for digital nets, see \cite[Eq. (3.27)]{skriganov} for ordered MDS codes). This bound is sharp since, simply adding a point at infinity, classical multiplicity codes \cite{nrt} already attain it.

The restriction $ 2 \leq k \leq r \ell m - 2 $ is necessary, since otherwise there exist MOSRD codes for any $ \ell $ and $ q $.

\begin{example} [\textbf{OSR repetition codes}]
Let $ \ell $ and $ q $ be arbitrary. Assume that $n_i\leq m$ for all $i\in[\ell]$. Let $ \bbeta_i \in \F_{q^m}^{n_i} $ with $ \F_q $-linearly independent components and define
$$ X_i = \left( \begin{array}{c}
\bbeta_i \\
\bbeta_i \\
 \vdots \\
\bbeta_i
\end{array} \right) \in \F_{q^m}^{r \times n_i}, $$
for $ i \in [\ell] $. Set $ X = (X_1,X_2, \ldots, X_\ell) \in \F_{q^m}^{r \times n} $ and let $ \C \subseteq \F_{q^m}^{r \times n} $ be the one-dimensional $ \F_{q^m} $-linear code generated by $ X $. Since $ \wt_R(X_i^{[j]}) = \wt_R(X_i^{(j)}) = n_i $, for all $ i \in [\ell] $ and $ j \in [r] $, then $ \dd_{SR}^+(\C) = \dd_{SR}^-(\C) = rn $. Thus $ \C $ is both a positive and negative $ \F_{q^m} $-linear MOSRD code of dimension $ 1 $. By Theorem \ref{th_mosrd_duality}, so is its dual $ \C^\perp $, of dimension $ rn - 1 $. 
\end{example}

Observe that, for the MSRD case ($ r=1 $) with $ m \geq 2 $, linearized Reed--Solomon codes \cite{linearizedRS} achieve the length $ \ell = q-1 $ and cannot always be extended for the sum-rank metric using square matrices, as shown in \cite[Prop. 11]{doubly}. Thus we make the following conjecture for linear codes (although we could similarly propose it for nonlinear codes).

\begin{conjecture} [\textbf{MOSRD}]
Assume $ m = n_1 = n_2 = \ldots = n_\ell $. If there exists an $ \F_{q^m} $-linear MOSRD code $ \C \subseteq \F_{q^m}^{r \times n} $ with $ k = \dim_{\F_{q^m}}(\C) $ and $ 2 \leq k \leq r \ell m - 2 $, then 
\begin{alignat}{2}
\ell &\leq q+1
&\qquad\qquad&
\text{if } m=1, \textrm{ and } k \notin \{ 3,\ell - 3 \} \textrm{ if } r=1 \textrm{ and } q \textrm{ even},\\
\ell &\leq q + 2
&&
\text{if } m=r=1, k \in \{ 3,\ell - 3 \} \textrm{ and } q \textrm{ even},\\
\ell &\leq q-1
&&
\text{if } m \geq 2, \textrm{ and } \min \{ k, \ell m - k \} \geq 4 \textrm{ if } r=1.
\end{alignat}
\end{conjecture}

Essentially, the first two constitute the classical MDS conjecture (and the already solved ordered MDS case $ r \geq 2 $), whereas the third one is the proper extension to MOSRD codes. Notice, however, that for the non-MSRD case ($ r \geq 2 $), the only case left to prove by Theorem \ref{th_bounds_ell_q} is the case with $ m = 2 $ and $ k $ odd. Thus the conjecture is mostly an MSRD conjecture. In the pure MSRD case ($ m \geq 2 $ but $ r=1 $), the conjectured bound was also recently proposed in \cite{astore}, where the authors include the restriction $ \min \{ k, \ell m - k \} \geq 4 $ since otherwise they provide constructions that exceed the bound.

Finally, the situation for general rectangular MOSRD codes (i.e., $ n_1=n_2= \ldots = n_\ell $ but not necessarily equal to $ m $) is significantly more complicated. Already in the MSRD case ($ r=1 $), in \cite{generalMSRD}, constructions of MSRD codes in this regime can achieve block lengths $ \ell $ that grow polynomially in $ q $ with degree much larger than $ 1 $. Some upper bounds on $ \ell $ can be found in \cite[Th. VI.12]{fundamental}, but we do not attempt to make any conjecture in this general case.

\section*{Disclosure on AI usage}

OpenAI's ChatGPT-6 was used to find related literature, for improving the Latex code and to find spelling, notational and mathematical errors in the manuscript. It also helped with some computations in Sections \ref{sec_weight}, \ref{sec_upper_bounds_m} and \ref{sec_upper_bounds_ell}. The mathematical ideas behind the definitions, statements and proofs are all due to the author, who takes responsibility for their correctness.

\section*{Acknowledgements}

This work has been partially supported by Grant PID2022-138906NB-C21 funded by MICIU/AEI/ 10.13039/501100011033 and by ERDF/EU, and Grant CLU-2025-1-02-IMUVA by the Department of Education of the Junta de Castilla y León and FEDER Funds.

{\small

\bibliographystyle{plain}

}

\newpage

\section*{Appendix: A worked example}

\textbf{AI disclosure:} Several computations in this worked example were performed with assistance from OpenAI's ChatGPT-6. In particular, the computation of the minimum distance in (\ref{eq_min_dist_app}) was entirely verified by this AI model, using Python with exact finite-field arithmetic. The author takes responsibility for the correctness of this Appendix. \\

Take $ q = 4 $, $ \ell = 3 $ and $ r = m = n_1 = n_2 = n_3 = 2 $. Thus $ n = 6 $ and the ambient space is
$$ \F_{q^m}^{r \times n} = \F_{16}^{2 \times 6} \cong (\F_{16}^{2 \times 2})^3. $$
Consider $ \F_4 = \F_2(\omega) $, where $ \omega^2 + \omega + 1 = 0 $, and $ \F_{16} = \F_4(\theta) $, where $ \theta^2 + \theta + \omega = 0 $. In particular, $ \theta^4 = \theta + 1 $ and $ \theta^5 = \omega $, thus $ \theta $ is a primitive element of $ \F_{16} $. We have the automorphism $ \sigma:\F_{16}\longrightarrow\F_{16} $ given by
$$ \sigma(a)=a^4 , $$
for all $ a \in \F_{16} $. Choose the three starting points in $ \F_{16}^* $
$$ a_1 = 1, \quad a_2 = \theta \quad \textrm{and} \quad a_3 = \theta^2. $$
They are pairwise nonconjugate, since (by Hilbert's Theorem 90), their norms over $ \F_4 $ are, respectively,
$$ N_{\F_{16}/\F_4}(a_1) = 1, \quad N_{\F_{16}/\F_4}(a_2) = \theta^5 = \omega, \quad \textrm{and} \quad N_{\F_{16}/\F_4}(a_3) = \theta^{10} = \omega^2 = \omega + 1. $$
For $ i = 1,2,3 $, if we choose $ \beta_i = \theta a_i = \theta^i \in \F_{16}^* $, then 
$$ {\rm Tr}_{\F_{16}/\F_4}(\beta_i a_i^{-1}) = {\rm Tr}_{\F_{16}/\F_4}(\theta) = \theta+\theta^4 = 1 \neq 0. $$
By Theorem \ref{th_char_multiplicity}, we obtain the multiplicity sequences
$$ \begin{aligned}
\mathbf a_{1,1}&=(1) \in \F_{16},
&
\mathbf a_{1,2}&=(1,\theta^3) \in \F_{16}^2,\\
\mathbf a_{2,1}&=(\theta) \in \F_{16},
&
\mathbf a_{2,2}&=(\theta,\theta^7) \in \F_{16}^2,\\
\mathbf a_{3,1}&=(\theta^2) \in \F_{16},
&
\mathbf a_{3,2}&=(\theta^2,\theta^{11}) \in \F_{16}^2.
\end{aligned} $$
The reader may verify by direct computation that each of the three skew polynomials
$$ \begin{aligned}
P_{\aaa_{1,2}} & = & (x-\theta^3)(x-1) & = & x^2 + \theta^{14} x + \theta^3 , \\
P_{\aaa_{2,2}} & = & (x-\theta^7)(x- \theta) & = & x^2 + \theta^3 x + \theta^8 , \\
P_{\aaa_{3,2}} & = & (x - \theta^{11})(x - \theta^2) & = & x^2 + \theta^7 x + \theta^{13},
\end{aligned} $$
only has one zero in $ \F_{16} $ on the right ($ 1 $, $ \theta $ and $ \theta^2 $, respectively), using arithmetic evaluations as in Definition \ref{def_ev_arithmetic}. We will choose the same ordered basis 
$$ \bbeta_i = (\beta_{i,1}, \beta_{i,2}) = (1, \theta) \in \F_{16}^2 $$
of $ \F_{16} $ over $ \F_4 $, for $ i = 1,2,3 $. Let
$$ \C = \M_4(\mathbf a,\boldsymbol\beta) = \{C(F):F\in\F_{16}[x;\sigma],\ \deg(F)<4\} \subseteq \F_{16}^{2 \times 6} . $$
A basis of $ \C $ as an $ \F_{16} $-linear code is given by
$$ C(1),\quad C(x),\quad C(x^2)\quad \textrm{and} \quad C(x^3). $$
From the definition of arithmetic Hasse derivatives (Definition \ref{def_hasse_arithmetic}), we see that $ D_{(a,b)}(1)=0 $, $ D_{(a,b)}(x)=1 $,  $ D_{(a,b)}(x^2) = b+\sigma(a) $ and $ D_{(a,b)}(x^3) = \sigma(b)b+aD_{(a,b)}(x^2) $, for a multiplicity sequence $ (a,b) \in \F_{16}^2 $. Therefore, after some tedious computations, we have that
\begin{equation*}
\begin{split}
C(1) & =
\left(
\begin{array}{cc|cc|cc}
1&\theta&1&\theta&1&\theta\\
0&0&0&0&0&0
\end{array}
\right), \\
C(x) & =
\left(
\begin{array}{cc|cc|cc}
1&\theta^4&\theta&\theta^5&\theta^2&\theta^6\\
1&\theta^4&1&\theta^4&1&\theta^4
\end{array}
\right), \\
C(x^2) & =
\left(
\begin{array}{cc|cc|cc}
1&\theta&\theta^5&\theta^6&\theta^{10}&\theta^{11}\\
\theta^{14}&1&\theta^3&\theta^4&\theta^7&\theta^8
\end{array}
\right), \\
C(x^3) & =
\left(
\begin{array}{cc|cc|cc}
1&\theta^4&\theta^6&\theta^{10}&\theta^{12}&\theta\\
\theta^3&\theta^7&\theta^8&\theta^{12}&\theta^{13}&\theta^2
\end{array}
\right).
\end{split}
\end{equation*}
Now, for a matrix $ X = (X_1,X_2,X_3) \in \F_{16}^{2 \times 6} $, with
$$ X_i = \left( \begin{array}{cc}
x_{i,11}&x_{i,12}\\
x_{i,21}&x_{i,22}
\end{array} \right) \in \F_{16}^{2 \times 2}, $$
for $ i = 1,2,3 $, we define its vectorization as
$$ \operatorname{vec}(X)
=
(x_{1,11},x_{1,12},x_{1,21},x_{1,22}
\mid
x_{2,11},x_{2,12},x_{2,21},x_{2,22}
\mid
x_{3,11},x_{3,12},x_{3,21},x_{3,22}) \in \F_{16}^{12} . $$
In this way, a generator matrix over $ \F_{16} $ of $ \C $, seen in $ \F_{16}^{12} $, would be
$$ G =
\left(
\begin{array}{cccc|cccc|cccc}
1&\theta&0&0
&
1&\theta&0&0
&
1&\theta&0&0
\\
1&\theta^4&1&\theta^4
&
\theta&\theta^5&1&\theta^4
&
\theta^2&\theta^6&1&\theta^4
\\
1&\theta&\theta^{14}&1
&
\theta^5&\theta^6&\theta^3&\theta^4
&
\theta^{10}&\theta^{11}&\theta^7&\theta^8
\\
1&\theta^4&\theta^3&\theta^7
&
\theta^6&\theta^{10}&\theta^8&\theta^{12}
&
\theta^{12}&\theta&\theta^{13}&\theta^2
\end{array}
\right) \in \F_{16}^{4 \times 12}. $$
As predicted by Theorem \ref{th_lin_mult_mosrd}, it can be verified that $ \dim_{\F_{16}}(\C) = 4 $ and 
\begin{equation}
\dd_{SR}^+(\C) = rn-k+1 = 2\cdot6-4+1 = 9.
\label{eq_min_dist_app}
\end{equation}
Therefore $ \C \subseteq \F_{16}^{2 \times 6} $ is an MOSRD code with parameters
$$ [r,m,\ell;n_1,n_2,n_3;k,d]_q = [2,2,3;2,2,2;4,9]_4. $$
Notice also that $ m = n_1 = n_2 = n_3 = 2 $ and $ \ell = 3 = q-1 $, thus $ \C $ attains the bounds (\ref{eq_bound_m_ell_geq_2}) and (\ref{eq_bound_mgeq3}). 

Finally, as predicted by Corollary \ref{cor_LMC_QMSRD}, $ \C $ has minimum sum-rank distance $ d_{SR}(\C) = 5 $ (with codewords seen as tuples in $ (\F_{4}^{4 \times 2})^3 $) and minimum Hamming distance $ d_H(\C) = 5 $ (where Hamming weights of codewords in $ \F_{4}^{4 \times 6} $ are their number of nonzero columns), which coincide with
$$ n - \frac{k}{r} + 1 = 6 - \frac{4}{2} + 1 = 5. $$
That is, $ \C $ is also MSRD and MDS, and since it is MOSRD, it is also ordered MDS.

\end{document}